\documentclass[iop]{emulateapj}

\usepackage{natbib}
\usepackage{amsmath}
\usepackage{hyperref}

\begin{document}

\title{Subsurface magnetic field and flow structure of simulated sunspots}
\shorttitle{Subsurface magnetic field and flow}

\author{Matthias Rempel}

\shortauthors{M. Rempel}
 
\affil{High Altitude Observatory,
  NCAR, P.O. Box 3000, Boulder, Colorado 80307, USA}

\email{rempel@ucar.edu}

\begin{abstract}
  We present a series of numerical sunspot models addressing the subsurface
  field and flow structure in up to $16$~Mm deep domains covering up to 
  2 days of temporal evolution. Changes in the photospheric
  appearance of the sunspots are driven by subsurface flows in several Mm
  depth. Most of magnetic field is pushed into a downflow vertex of the 
  subsurface convection pattern, while some fraction of the flux separates
  from the main trunk of the spot. Flux separation in deeper layers is 
  accompanied in the photosphere with light bridge formation in the
  early stages and formation of pores separating from the spot at later
  stages. Over a time scale of less than a day we see the development of a 
  large scale flow pattern surrounding the sunspots, which is dominated by 
  a radial outflow reaching about $50\%$ of the convective rms velocity in 
  amplitude. Several components of the large scale flow are found to be 
  independent from the presence of a penumbra and the associated Evershed flow.
  While the simulated sunspots lead to blockage of heat flux in the near 
  surface layers, we do not see compelling evidence for a brightness 
  enhancement in their periphery. We further demonstrate that 
  the influence of the bottom boundary condition on the stability and long-term 
  evolution of the sunspot is significantly reduced in a $16$~Mm deep domain 
  compared to the shallower domains considered previously.
\end{abstract}


\received{}
\accepted{}

\maketitle

\section{Introduction}
The subsurface structure of sunspots has been of subject of theoretical
investigations for several decades. The two possible  extremes of magnetic 
configurations were discussed by \citet{Parker:1979b}: a monolithic
configuration vs. a clusters of individual flux tubes that is kept
together by converging flows in a suitable depth. On the observational
side evidence is inconclusive. Direct helioseismic measurements of the
magnetic field structure with an accuracy to determine the differences
between monolithic and cluster models are currently out of reach. More
promising are measurements of the subsurface flow structure, but also
there results are inconclusive. Time distance inversions by  
\citet{Zhao:etal:2001,Zhao:etal:2010} point toward inflows around sunspots
in an intermediate depth range from about $1.5$ to $5$~Mm (and corresponding
downflows underneath sunspots), which would be consistent with the 
expectations from a cluster model. On the other hand recent inversions
presented by \citet{Gizon:etal:2009,Gizon:etal:2010:err} show outflows in the
upper most $4.5$~Mm.

At photospheric levels most sunspots are surrounded by large scale outflows
(called ``moat flows'') with amplitudes of a few $100$~ms$^{-1}$ that were 
first found through 
tracking of magnetic features \citep{Sheeley:1969,Harvey:Harvey:1973},
Doppler measurements \citep{Sheeley:1972} and later helioseismic
measurements \citep{Gizon:etal:2000}. Several recent investigations focused
on possible connections between the Evershed flow and moat flow
\citep[see e.g.][]{SDalda:MPillet:2005,Cabrera-Solana:etal:2006,
VDominguez:etal:2008,Zuccarello:etal:2009,VDominguez:etal:2010},
so far the observational evidence is not clear enough to either proof 
or disproof a connection.

The subsurface field and flow structure has been also addressed through
means of numerical models. 2D axisymmetric models by 
\citet{Hurlburt:Rucklidge:2000}, \citet{Botha:etal:2006}, and 
\citet{Botha:etal:2008} show large scale flow patterns around sunspots.
The typical result is a converging collar flow in the proximity of the
spot and a diverging flow further out, a situations which is
essentially in agreement with the cluster model as well as the inversions
by \citet{Zhao:etal:2001,Zhao:etal:2010}. Recently \citet{Botha:etal:2011}
expanded this work to 3D and found comparable results with regard to the 
axisymmetric flow components. Differences occurred with regard to the
process of sunspot decay: the azimuthal cell structure allows for flux to
escape from the central flux concentration even if the average flow is
converging. At this point neither the 2D axisymmetric nor the 3D simulations
contain a penumbra and the connection of the larger scale deep seated flows
to photospheric flows remains an open question.
 
Over the past five years there has been a substantial progress in
3D MHD models that include a realistic equation of state and radiative
transfer. Owing to the wide range of length and time scales involved
in sunspot structure and evolution it is currently not possible to address all 
aspects of sunspot structure and evolution in a single numerical simulation.
The formation, evolution and decay of pore-size flux concentrations has been 
modeled by \citet{Bercik:etal:2003,Cameron:etal:2007b,
Kitiashvili:etal:2010:pore}, 
typically resulting in converging flows in the proximity of the pore.
The focus of recent numerical models such as \citet{Schuessler:Voegler:2006},
\citet{Heinemann:etal:2007}, \citet{Rempel:etal:2009}, and 
\citet{Kitiashvili:etal:2009} was primarily the sunspot fine structure and 
origin of the Evershed flow. To this end those models focused on smaller 
subsections and rather short temporal evolution of a few hours. The
simulations by 
\citet{Rempel:etal:Science,Rempel:2010:IAU,Rempel:2011} were the 
first MHD simulations with a sufficient domain size in the horizontal direction
to capture complete sunspots, while still resolving sunspot fine structure.
However, the vertical domain size of $6.144$~Mm remained still too shallow and 
the overall time evolution of up to 6 hours too short to properly address the 
subsurface structure of the sunspot as well as development of moat flows.
While these simulations present a unified picture for the magneto convective
origin of sunspot fine structure they leave several fundamental aspects
with regard to the subsurface structure unanswered. 

Evidence for a larger scale diverging flow is already present in the 
simulations 
of \citet{Heinemann:etal:2007}, \citet{Rempel:etal:2009},  
\citet{Rempel:etal:Science}, and \citet{Rempel:2011}, but the overall 
development of this flow 
pattern was heavily influenced by either the domain size or the rather 
short time span covered. We will relax most of these constraints in this 
investigation by focusing on numerical simulations with lower resolution, 
but deeper domains (up to 16 Mm) and much longer temporal evolution 
(up to 2 days). Simulations in deeper domains are less influenced by the 
bottom boundary condition and allow us to study processes 
related to the fragmentation of sunspots and development of large
scale flow patterns. 

In Sect. \ref{sect:models} we describe the setup of the numerical 
simulations used in this investigation. Sect. \ref{sect:results} presents
the time evolution and structure of the subsurface magnetic field and
flow structure. The influence of the (unfortunately unavoidable) bottom 
boundary condition is analyzed in Sect. \ref{sect:bottom-bnd}, the
findings are discussed in Sect. \ref{sect:discussion} and summarized in
Sect. \ref{sect:conclusions}.

\section{Numerical models}
\label{sect:models}
The details of the numerical model and underlying physics are described in
\citet{Voegler:etal:2005} and \citet{Rempel:etal:2009}. Models presented here
use a setup similar to those of \citet{Rempel:etal:Science}, but differ in
domain size, resolution as well as boundary conditions.

We report here on a series of numerical simulations which were performed in
a computational domain with a horizontal extent of $49.152$~Mm and a vertical 
extent of $16.384$~Mm. We used a rather low resolution of $96$~km horizontally 
and $32$~km vertically to allow for long simulation runs covering
up to 2 days of solar time. The simulations were started from a
non-magnetic convection simulation, which was evolved for $42$ hours to 
allow for thermal relaxation
of the stratification. We then inserted into the domain an axisymmetric,
self-similar sunspot with $1.2\cdot 10^{22}$~Mx flux and a field strength of
$16$~kG at the bottom of the domain. While the domain is periodic in the
horizontal direction, the top boundary is closed for the vertical mass flux
and slip free for horizontal motions. The magnetic field is matched to a 
potential field extrapolation. At the bottom boundary the magnetic field is
vertical and we consider here two formulations with regard to flows. The 
first boundary condition sets all velocity 
components to zero in regions where the field strength exceeds 5 kG, while it
allows for convective motions to cross the bottom boundary everywhere else:
in outflow regions all velocity gradients are set to zero and thermodynamic 
variables are extrapolated into the ghost cells while inflows have purely 
vertical
velocities and a prescribed entropy to maintain the solar energy flux.   
The second boundary condition is an open boundary everywhere in the domain
regardless of the magnetic field strength. In contrast to the first boundary
condition we set the gradient of all mass flux components to zero, allowing
also for horizontal flows in inflow regions (a zero gradient of the
horizontal flow velocity instead of mass flux turns out to be unstable on
long time scales). While the first boundary condition strongly suppresses
the decay of the sunspot by limiting the motions at the foot point, the second
boundary allows for horizontal exchange between magnetized and unmagnetized
regions as well as vertical flows changing the mass content of the sunspot.
Note that the first boundary condition does not allow for heat flux entering
the domain in strong field regions, while in the second case heat exchange
is not prohibited by the boundary, although in general still strongly 
suppressed by the magnetic field. In both cases the energy radiated away in 
the sunspot umbra comes mostly from the heat content of the stratification, 
i.e., the umbra and stratification underneath slowly cool down as the simulation
progresses. The most dramatic temperature changes happen during the first few
hours of the simulations, later most of the energy comes from several Mm deep
layers, where the heat capacity is sufficiently large to prevent strong 
temperature changes. A similar behavior is also found in simpler models
such as \citet{Schuessler:Rempel:2005}.

In the following discussion we focus first on the simulation with the
closed boundary condition in strong field regions, which leads to an overall
more stable sunspot. We contrast these results with simulations using the 
second boundary condition in Sect. \ref{sect:bottom-bnd}.

We present in addition results from a simulation in a $73.728$~Mm wide and
$9.216$~Mm deep domain at a resolution of $48$~km horizontally and $24$~km
vertically, which was evolved for a 
total of $24$ hours. In this simulation we use a different top boundary 
condition, which enforces a more inclined magnetic field than the previous 
cases using a potential field. To this end we enhanced the horizontal field
components by a factor of 2 compared to a potential field extrapolation. 
Together with 
the higher resolution this boundary leads to a stable extended penumbra with
average Evershed flows of about $4\,\mbox{km\,s}^{-1}$. The bottom boundary 
is closed in 
the inner most $8$~Mm to prevent decay of the sunspot. The total flux of this
sunspot is again $1.2\cdot 10^{22}$~Mx, the field strength at the bottom 
boundary is $10$~kG. We will present results from this simulation with 
regard to axisymmetric mean flows and compare them to the previously
mentioned simulation runs to investigate the influence of penumbra
and Evershed flow on large scale flows surrounding sunspots. For
a more detailed description of this simulation run and a series of 
simulations with the different top boundary condition we refer to a
separate forthcoming publication.

\section{Results}
\label{sect:results}
\begin{figure*}
  \centering 
  \resizebox{0.9\hsize}{!}{\includegraphics{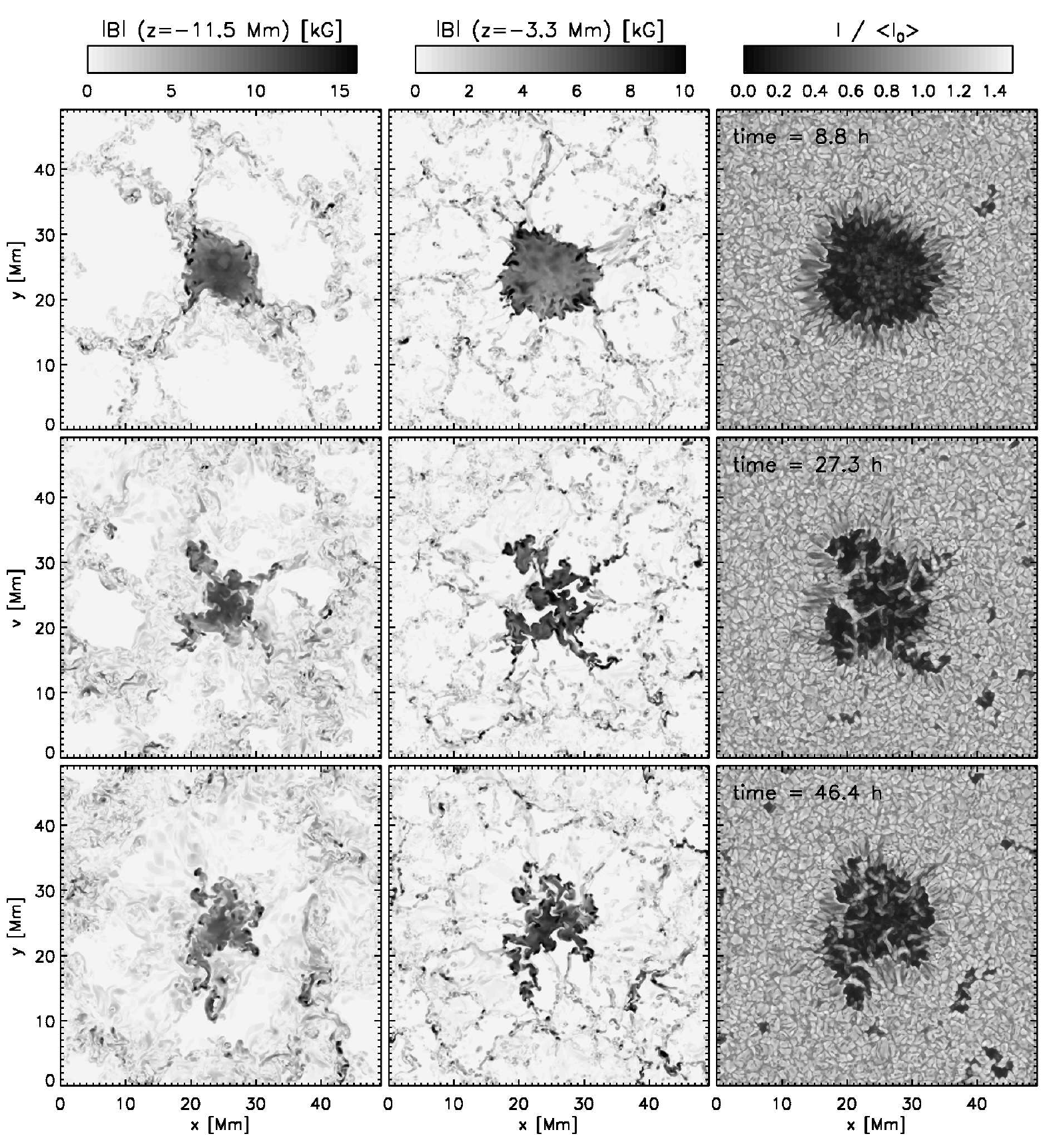}}
  \caption{Temporal evolution of subsurface magnetic field strength at
    two depth levels (left column: $z=-11.5$~Mm, middle column: $z=-3.3$~Mm) 
    and surface appearance of sunspot (right column: intensity). Top to bottom
    snapshots at $t=8.8$ hours, $t=27.3$ hours and $t=46.4$ hours are shown.
    The magnetic field is
    swept into a downflow vertex of the subsurface convection pattern.
    A fraction of the initial magnetic flux is separating from the main 
    trunk along downflow lanes. At the surface the separation of flux is
    accompanied with the formation of light bridges and pores
    surrounding the spot at later stages. An animation is available in the 
    online material.
    \label{fig:f1}
  }
\end{figure*}

\begin{figure*}
  \centering 
  \resizebox{0.9\hsize}{!}{\includegraphics{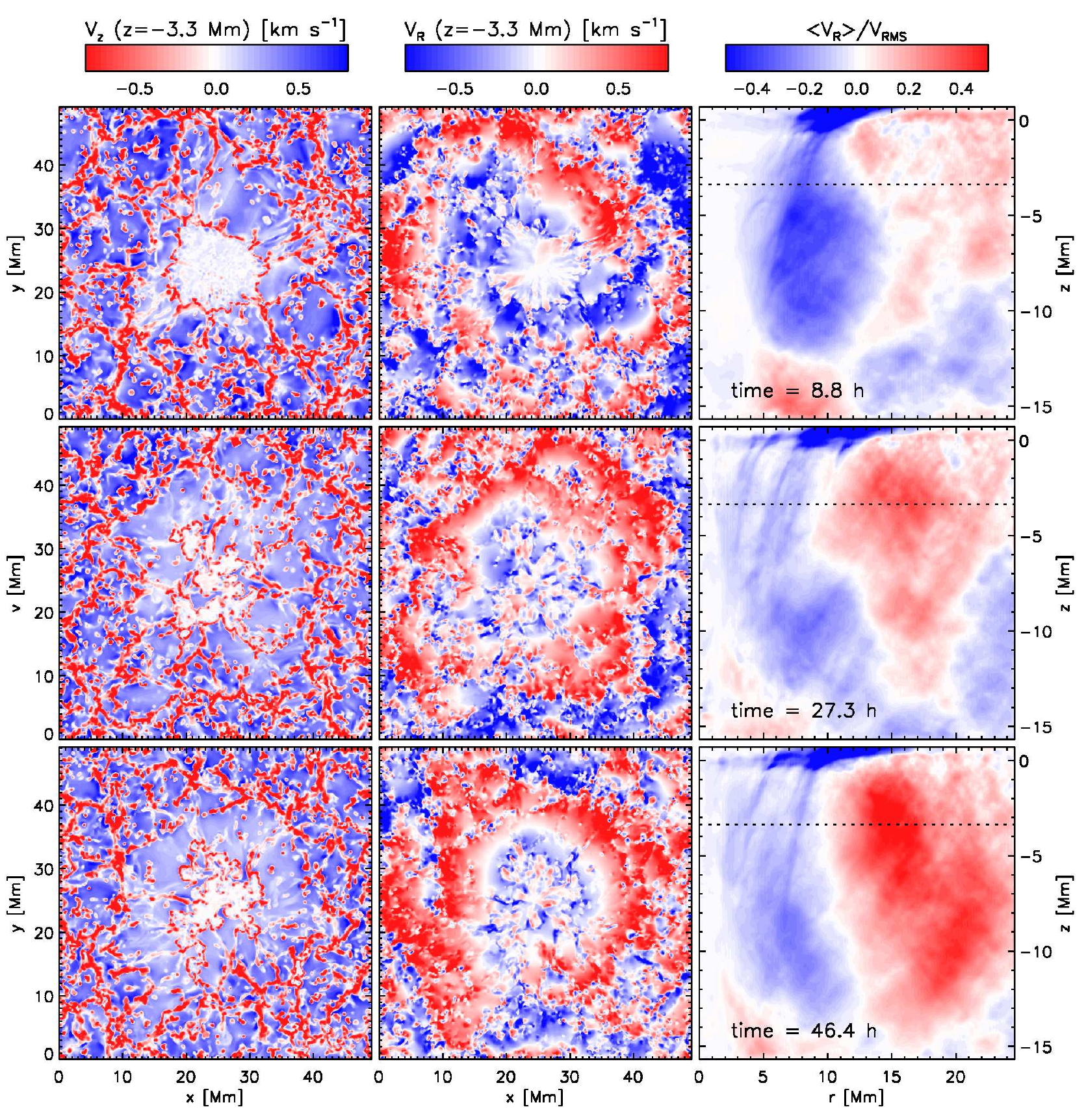}}
  \caption{Evolution of large scale flows shown for the same snapshots
    presented in Figure \ref{fig:f1}. The left column present vertical
    velocity at a depth of $z=-3.3$~Mm, blue colors indicate upflows.
    The middle column presents radial flow velocity at a depth of $z=-3.3$
    Mm, red colors indicate outflows. The right column presents azimuthal
    averages of the radial flow velocity, normalized by the convective rms
    velocity. The dotted horizontal line indicates the depth level of
    $z=-3.3$~Mm used in the left and middle column. Convection cells
    arrange in a ring-like fashion around the sunspot. As a consequence
    horizontal convective flows lead to a mean flow reaching about
    $50\%$ of the convective rms velocity during the later stages of the 
    simulation. \label{fig:f2}
  }
\end{figure*}

\begin{figure*}
  \centering 
  \resizebox{0.95\hsize}{!}{\includegraphics{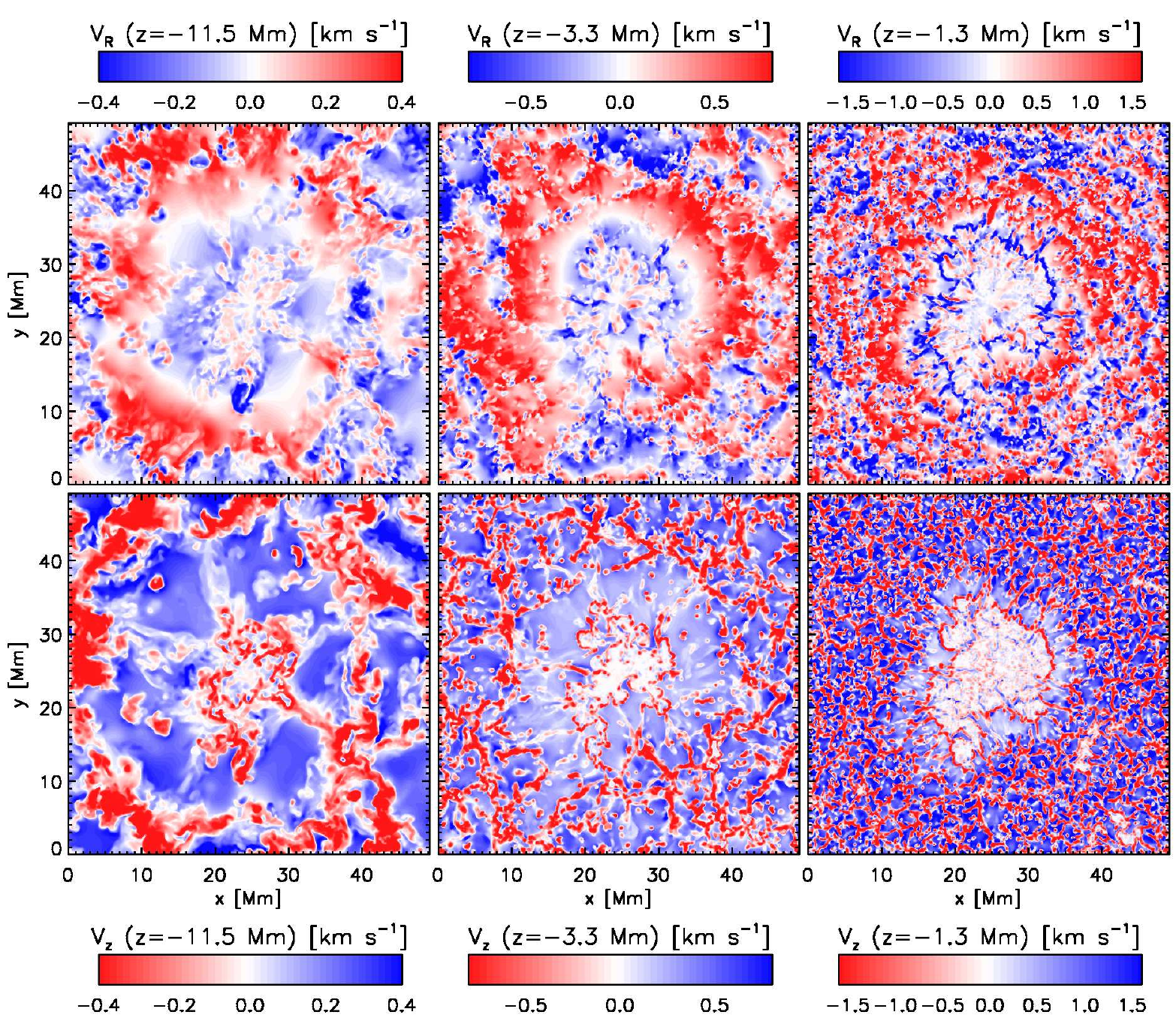}}
  \caption{Radial and vertical flow velocities at 3 depth levels for
    the snapshot at $t=46.4$ hours. The top row shows radial, the bottom row
    vertical flow velocity. Left to right the depth levels $z=-11.5$~Mm,
    $z=-3.3$~Mm, and $z=-1.3$~Mm. The ring-like arrangement of convection
    cells is present at all depth levels, the diameter of the region in which
    the presence of the sunspot modifies the convection pattern is
    increasing with depth as the intrinsic scale of convection is increasing
    with pressure scale height. An animation is available
    in the online material. \label{fig:f3}
  }
\end{figure*}

\subsection{Temporal evolution of subsurface magnetic field}
\label{sect:field}
Figure \ref{fig:f1} highlights the connection between the temporal evolution
of the subsurface magnetic field and appearance of the sunspot in the 
photosphere. The corresponding subsurface flow structure is presented
in Figure \ref{fig:f2} and discussed further in section \ref{sect:flow}.
The left and middle column show the magnetic field strength at
a depth of $11.5$ and $3.3$~Mm beneath the $\tau=1$ level of the plage
region surrounding the sunspot. The right column presents the surface intensity
(gray intensity for vertical direction). We show top to bottom
3 time steps about 18 - 19 hours apart. The sunspots starts as a circular 
sunspot due to our initial condition, but is changing its shape substantially
throughout the simulated time span of about 2 days. Several Mm beneath the 
photosphere most of the magnetic flux is collected into a downflow vertex
(in about $10$~Mm depth the intrinsic scale of convection becomes larger than
the diameter of the sunspot at that depth), while some of the magnetic flux 
separates from the main trunk along downflow lanes connecting to the vertex.
A qualitatively similar picture of sunspot decay was also found in the
recent 3D simulation of \citet{Botha:etal:2011}.
The flux separation driven by convective motions in deeper layers becomes 
manifest in the photosphere as light bridges in the earlier stages and pores
separating from the main spot during the later stages. While light bridges
have a photospheric appearance very similar to penumbral filaments, their
origin is of fundamentally different nature. As shown by 
\citet{Schuessler:Voegler:2006}, \citet{Rempel:etal:2009} and 
\citet{Rempel:etal:Science} umbral dots and penumbral filaments originate from
magneto-convection in strong magnetic field. The observed reduction of field
strength in these features is a consequence of overturning convection and does
not require the intrusion of ``field free'' plasma from beneath or outside the
sunspot. The magneto-convective motions responsible for these structures are
concentrated in the upper most $500$~km beneath the photosphere. In contrast 
to this light-bridges are the consequence of almost field free plasma entering
from beneath and the associated structure is deep reaching (several Mm).
A good example is the snapshot at $t=27.3$ hours shown in Figure \ref{fig:f1}.
The light-bridge entering the sunspot from the left side is clearly visible 
as an intrusion of field free plasma in $3.3$~Mm depth and even the field in 
$11.5$~Mm depth shows a similar signature. While narrow light bridges show 
the formation of a dark lane in their center (above the central upflow), wider 
light bridges break down into several granulation like convection cells. Good
examples for the latter are present starting from $t=39$ hours in the animation
provided for Figure \ref{fig:f1} in the online material. Light bridges with 
similar properties were also found in the simulations of 
\citet{Cheung:etal:2010}.

As a byproduct of the sunspot decay we find several pores surrounding the
sunspot. The pores show a wide spread in life times and we see a very strong
indication that the depth extent of the subsurface magnetic field structure
is the most relevant factor determining their temporal evolution. Pores such as
those found in the lower right corner of the snapshot at $t=46.4$ hours in
Figure \ref{fig:f1} are connected to layers deeper than $10$~Mm and are visible
in the photosphere for almost 24 hours. 

\begin{figure*}
  \centering 
  \resizebox{0.95\hsize}{!}{\includegraphics{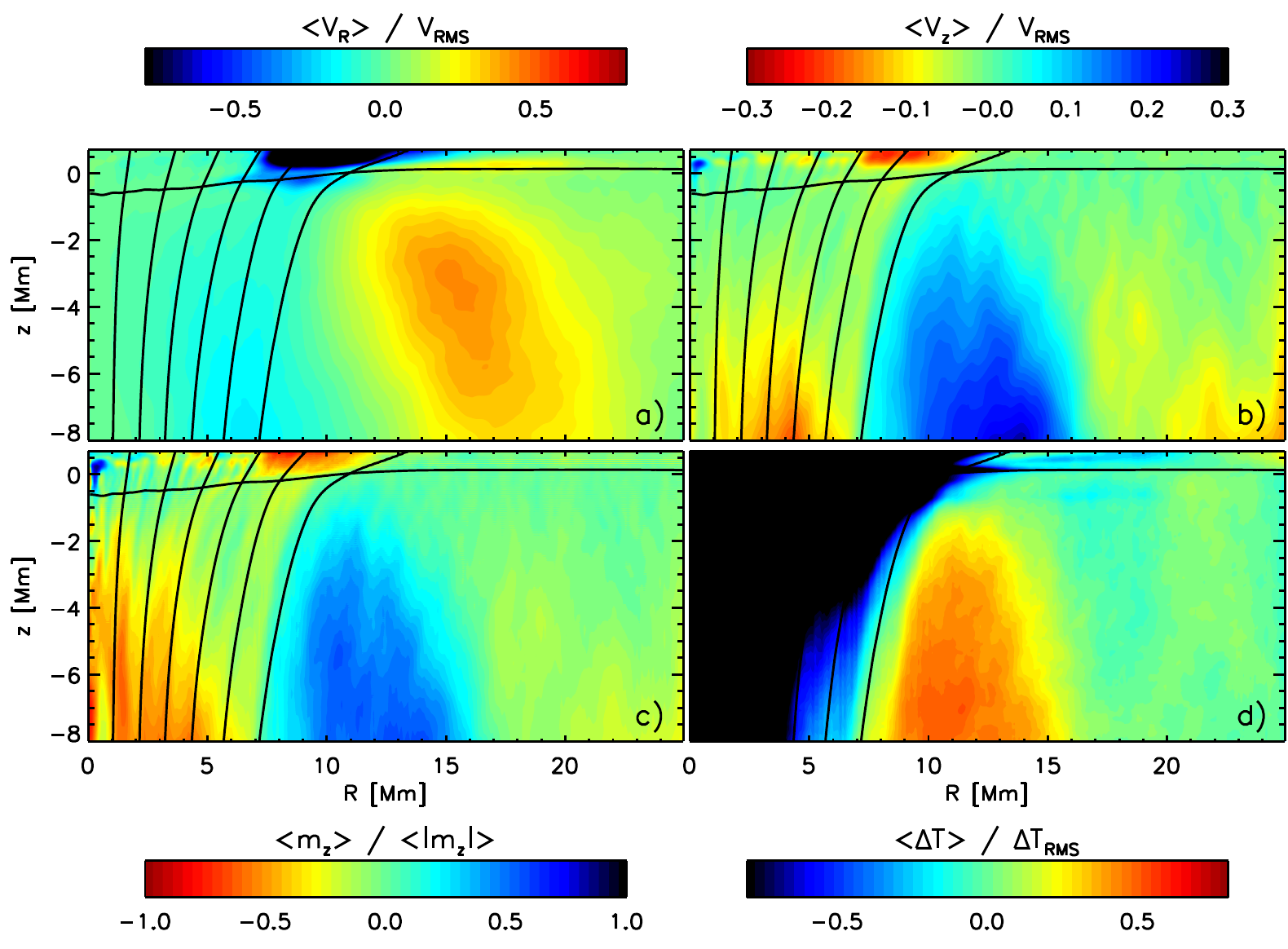}}
  \caption{Properties of the azimuthally and temporally ($15$ hours from 
    $t=35$ hours to $t=50$ hours) averaged flow in the simulation run without 
    penumbra. 
    Displayed are a) radial and b) vertical flow velocity relative to the 
    convective rms velocity (positive values indicate out/up-flows). 
    Panel c) presents the fraction of the mass flux present in the
    azimuthal mean component, panel d) the temperature 
    perturbation relative to the mean stratification computed outside 
    $R=25$~Mm (i.e. corners of the domain) in units of the rms fluctuation
    (outside $R=25$~Mm). It is evident that the main upflow coincides with 
    an annulus of enhanced temperature between $R=8$ and $R=16$~Mm. The
    thermal signature of the large flow is comparable to that of convective
    flows. We indicated the field lines of the azimuthally averaged magnetic
    field for reference, the almost horizontal line near the top indicates the
    $\tau=0.1$ level.
  }
  \label{fig:f4}
\end{figure*}

\begin{figure*}
  \centering 
  \resizebox{0.95\hsize}{!}{\includegraphics{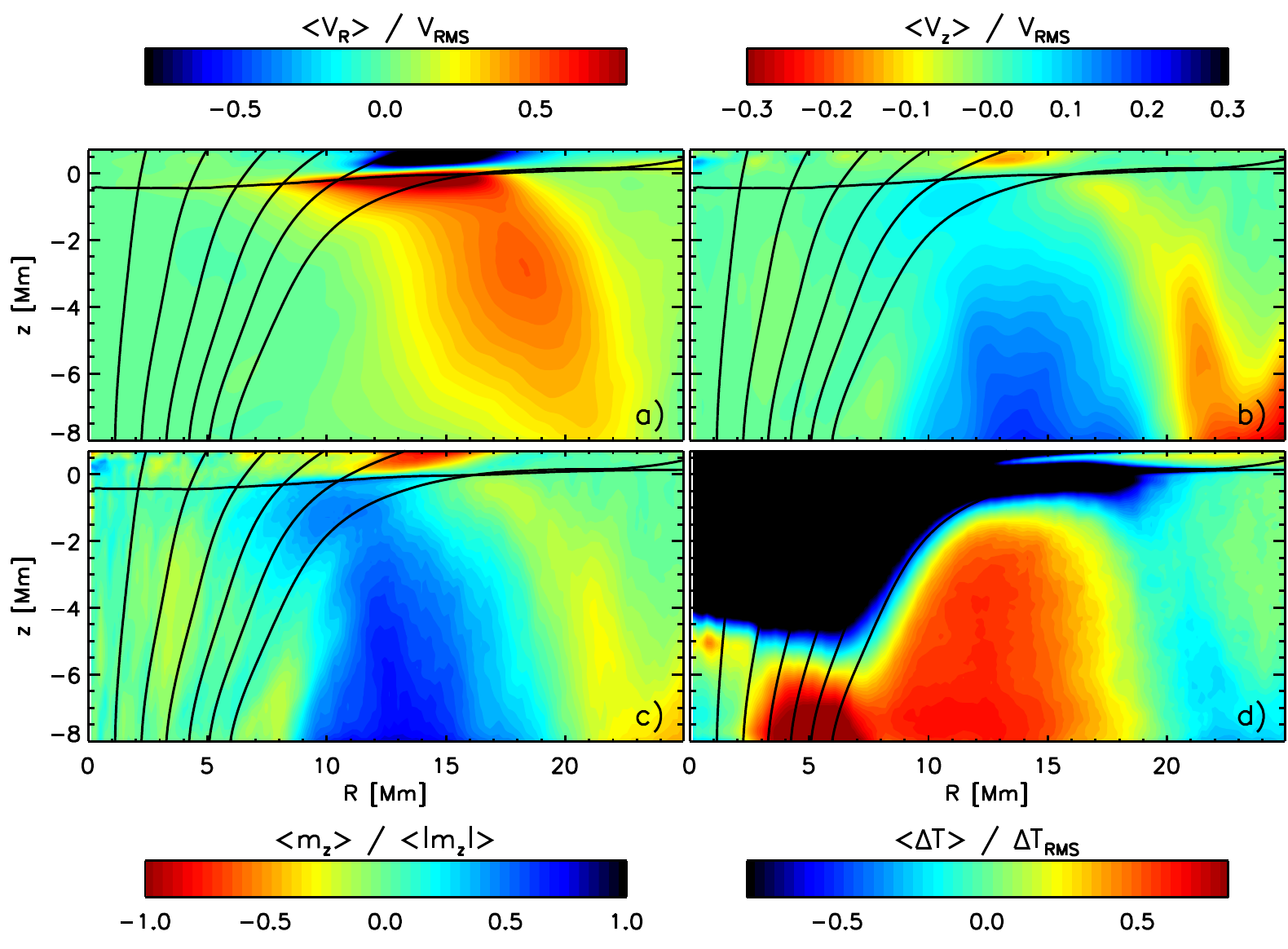}}
  \caption{Same quantities as in Figure \ref{fig:f4} for the simulation
    with penumbra and Evershed flow (see Figure \ref{fig:f10} for an intensity
    snapshot and magnetogram from this simulation run). Differences compared 
    to the previous
    case are restricted mostly to photospheric layers (see indicated $\tau=0.1$ 
    level). While the previous
    case shows an inflow for $R<10$~Mm and an outflow further out, radial
    flows are outward directed for all distances in the presence of a penumbra.
    The large scale outward directed 
    flow with amplitudes around $0.5 v_{\rm rms}$ and the associated thermal
    perturbation is very similar to the previous case. Overall this points
    toward a large degree of independence of this flow component from the
    presence of a penumbra and Evershed flow. \label{fig:f5}
  }
\end{figure*}

\begin{figure*}
  \centering 
  \resizebox{0.95\hsize}{!}{\includegraphics{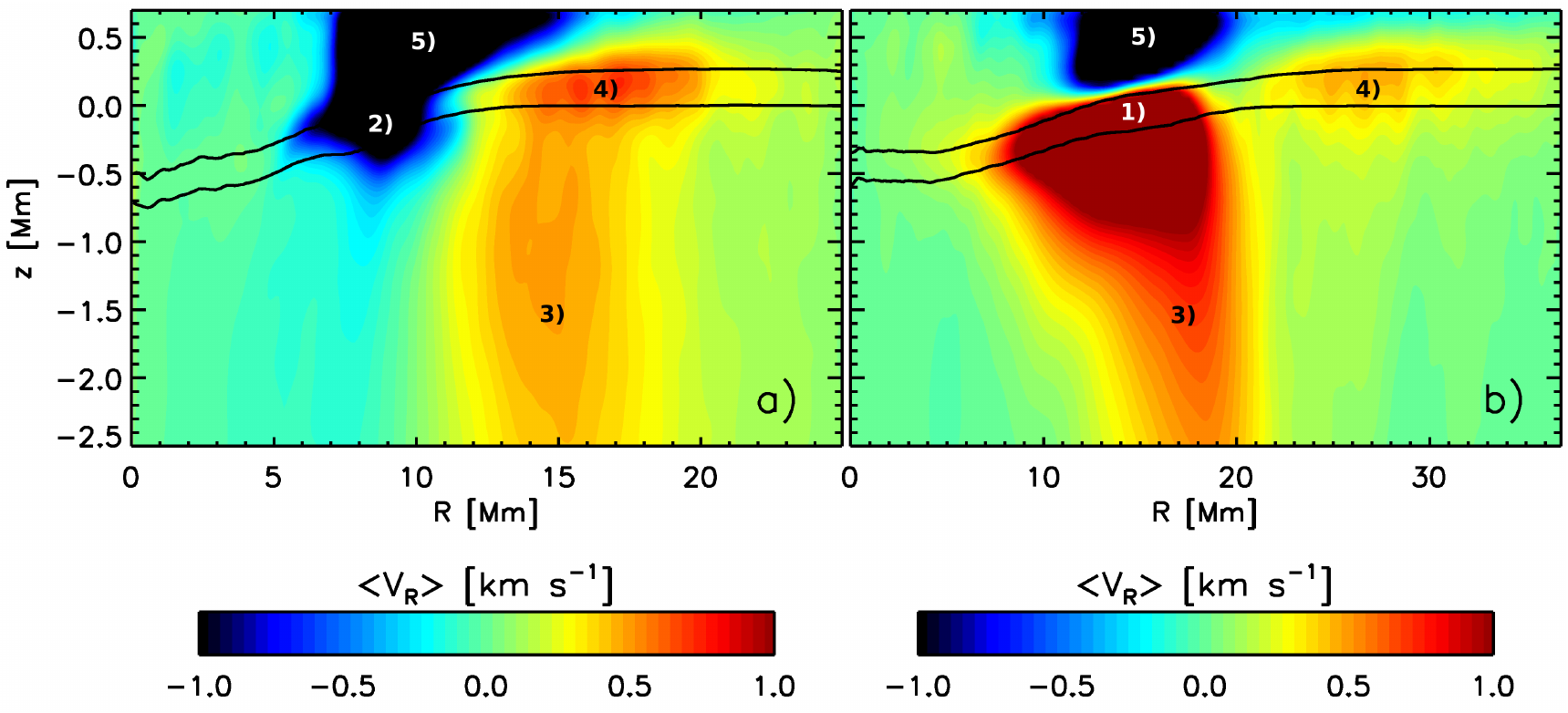}}
  \caption{Comparison of the radial flow amplitudes in the upper most
    $3$~Mm of the domain. Panel a) corresponds to the simulation
    presented in Figure \ref{fig:f4}, panel b) to Figure \ref{fig:f5}.
    In contrast to Figure \ref{fig:f4} and Figure \ref{fig:f5} the velocity is
    not normalized by the rms velocity. Also note the different horizontal
    scales and resulting aspect ratio in both panels. The two dark lines 
    indicate the $\tau=1$ and $\tau=0.01$ levels in both panels. In both 
    cases the flow amplitudes of the outflows surrounding the sunspots peak 
    above $\tau=1$, typical flow
    velocities are about $600\,\mbox{m\,s}^{-1}$ for the case without 
    penumbra and $400\,\mbox{m\,s}^{-1}$ for the case with penumbra.
    The numbers indicate flow components that are discussed further in the
    text.\label{fig:f6}
  }
\end{figure*}

\begin{figure}
  \centering 
  \resizebox{0.95\hsize}{!}{\includegraphics{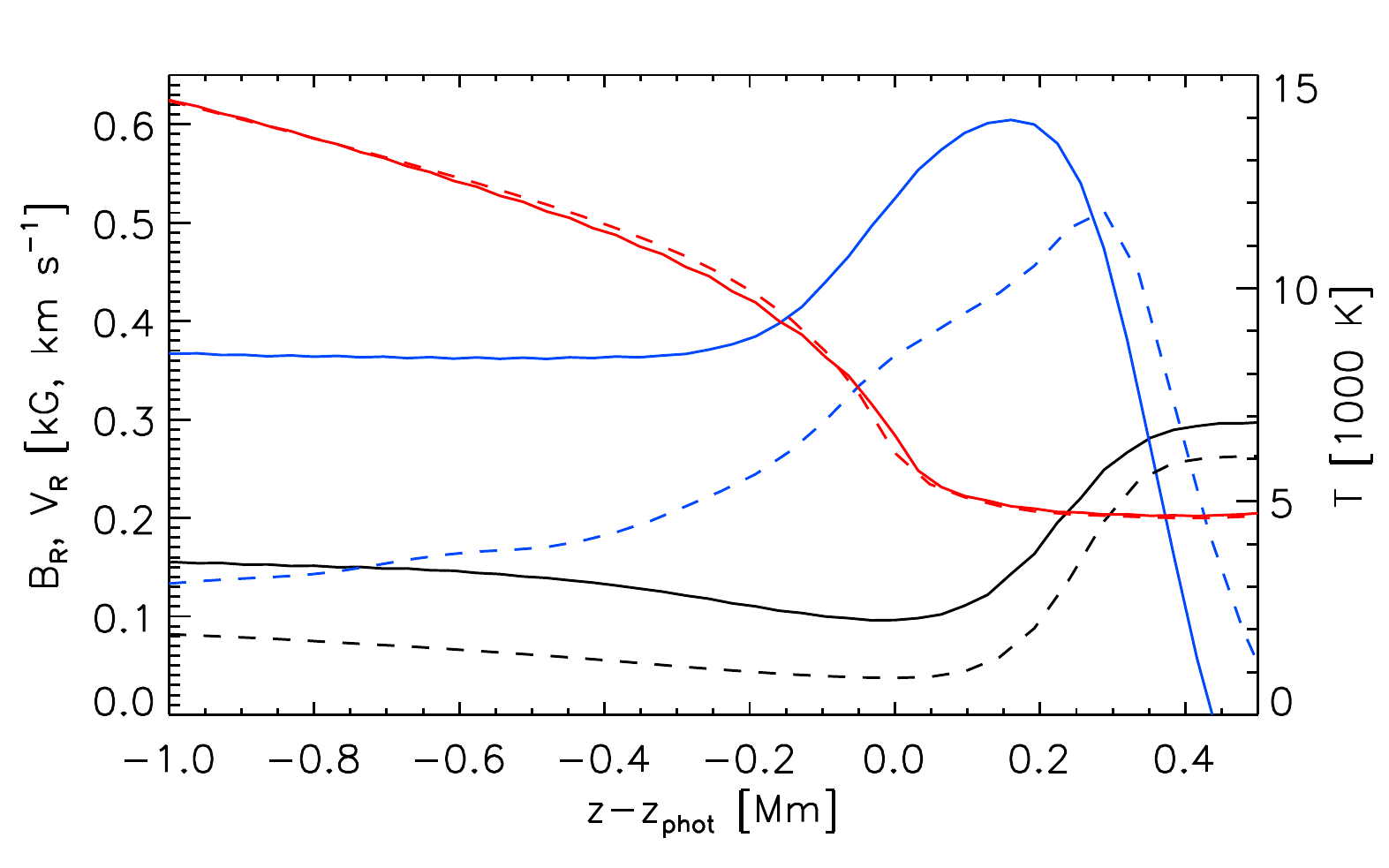}}
  \caption{
    Vertical profiles of the radial magnetic field strength (black),
    radial flow velocity (blue) and temperature (red). Solid lines show
    averages in between R=13 and R=20~Mm in Figure \ref{fig:f6}a, dashed
    lines show averages in between R=23 and R=28~Mm in Figure \ref{fig:f6}b.
    The fastest outflow velocities are found about 200~km above the $\tau=1$
    level, where the radial magnetic field strength increases and forms a
    magnetic canopy overlying the plage region surrounding the sunspots.
    \label{fig:f7}
  }
\end{figure}

\begin{figure*}
  \centering 
  \resizebox{0.95\hsize}{!}{\includegraphics{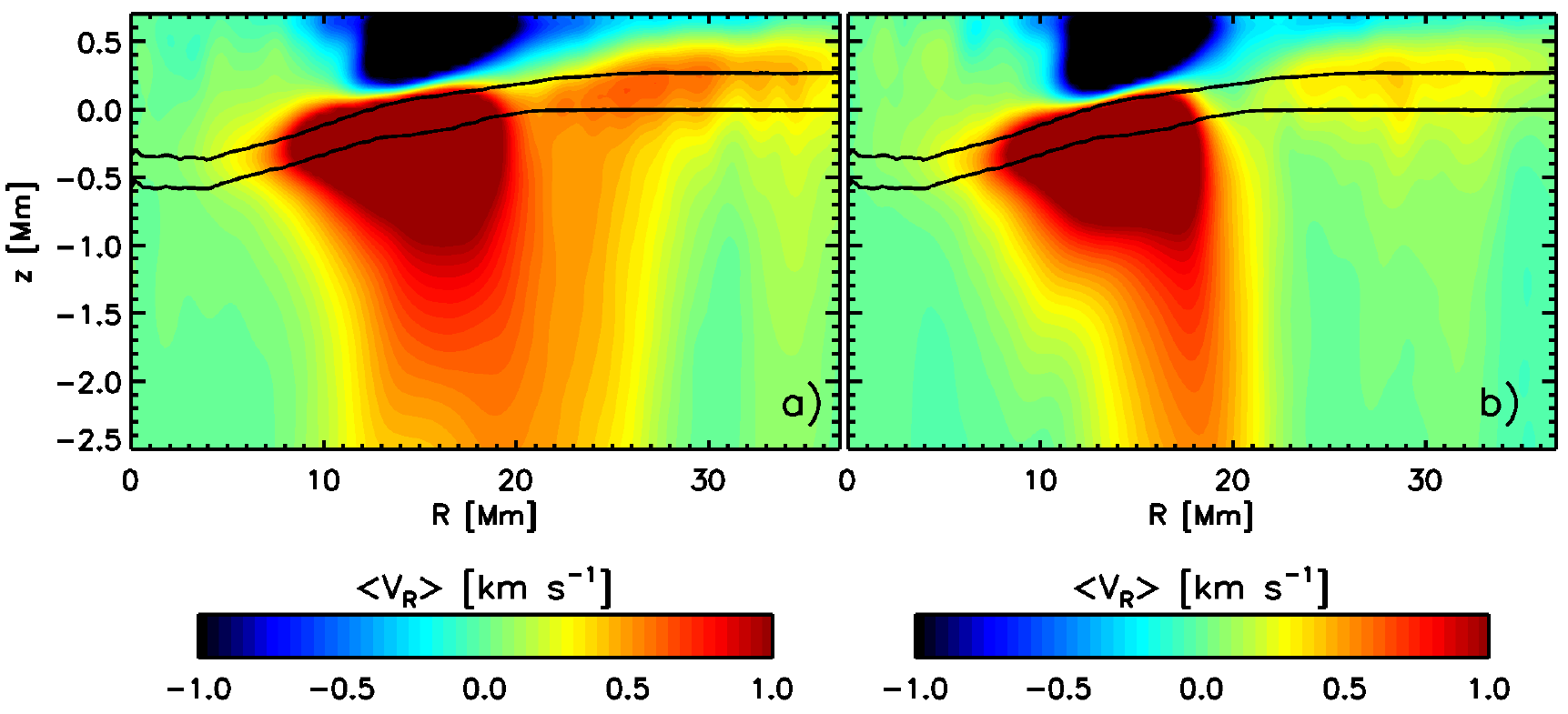}}
  \caption{Temporal evolution of large scale flows around the
    sunspot with penumbra. The quantities shown are similar to Figure 
    \ref{fig:f6}. In contrast to Figure \ref{fig:f6}b) where we presented a 
    time average
    from 12-24 hours, we show here in panel a) a time average from 6-12 
    hours and in panel b) the average from 18-24 hours. Overall we see a trend
    of decreasing flow velocities in the region $R>20$~Mm. While the
    photospheric flow speeds in panel a) are comparable to those found for
    the sunspot without penumbra (Figure \ref{fig:f6}a), panel b) shows a
    reduction by about a factor of 2. In addition the separation of the
    flow component 4 (as defined in Figure \ref{fig:f6}) becomes more evident.
    \label{fig:f8}
  }
\end{figure*}

\begin{figure*}
  \centering 
  \resizebox{0.95\hsize}{!}{\includegraphics{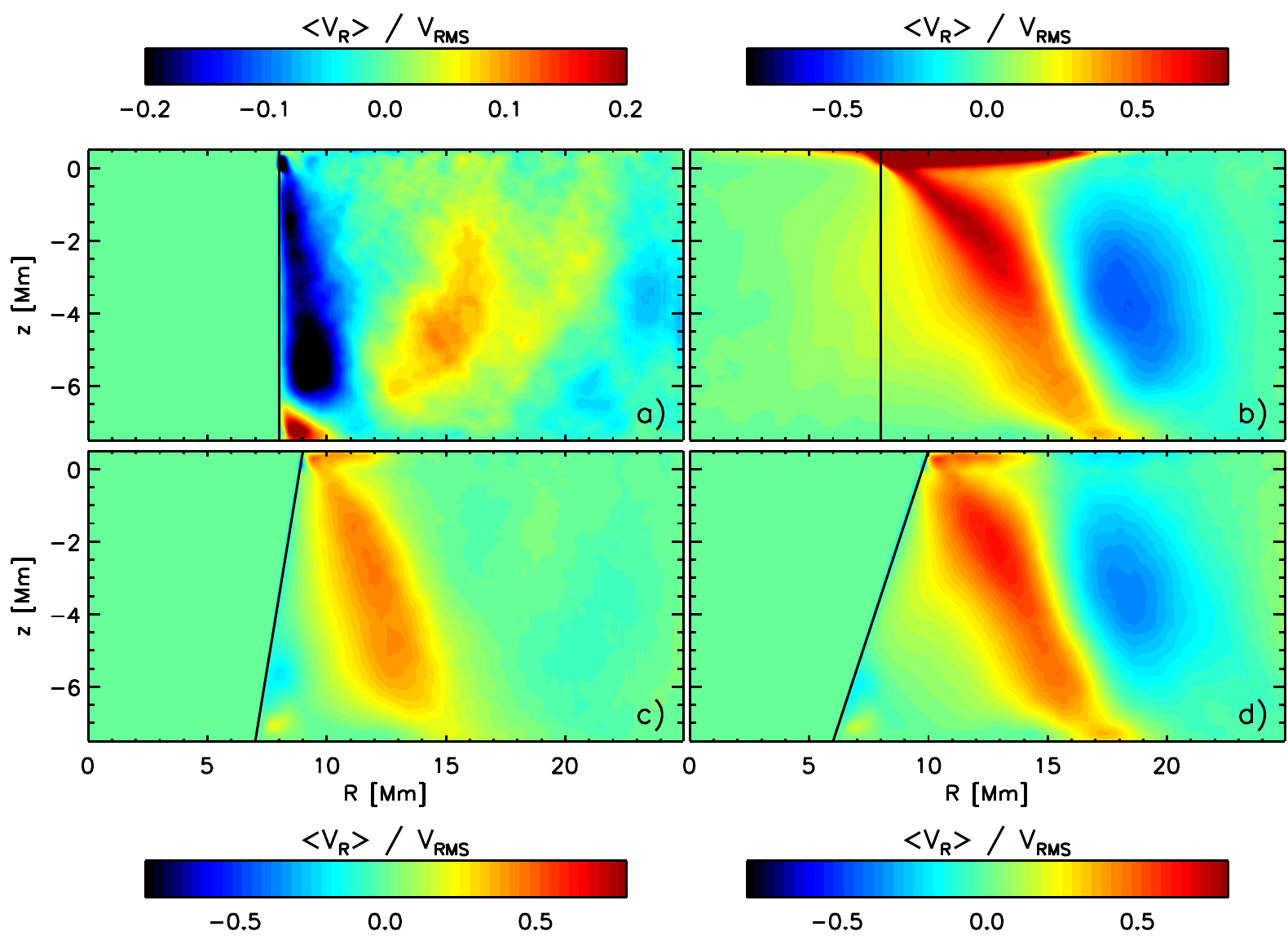}}
  \caption{Comparison of large scale flow patterns resulting from
    a combination  of geometric constraints and blockage of heat flux in 
    the photosphere. Panel a) shows the flow pattern developing
    around a cylinder with $8$~Mm radius due to the imposed geometric 
    constraint. Panel b) shows the flow patterns developing 
    in response to switching off the radiative cooling in the region
    $R<8$~Mm. Panels c) and d)  show flows around cone
    shaped obstacles with different opening angles, which result from a 
    combination of geometric and thermal effects.\label{fig:f9}
  }
\end{figure*}

\subsection{Temporal evolution of subsurface velocity field}
\label{sect:flow}
Recently \citet{Rempel:etal:Science} presented a numerical simulation
including an extended penumbra and Evershed flows with average flow amplitudes
faster than $4\,\mbox{km\,s}^{-1}$. Requirements for obtaining these 
results where a 
sufficiently high resolution to resolve penumbral filaments and the proximity
of an opposite polarity spot leading to more horizontal magnetic field. In the 
simulations we discuss here we focus on an individual sunspot and our 
horizontal grid resolution is 96 km instead of 32 km. Under these circumstances
we do not obtain a penumbra and near surface flows are converging toward the 
spot with a velocity of several $\mbox{km\,s}^{-1}$. A 
similar behavior is observed near pores, where the granules at the periphery 
of the umbra lead to a converging flow pattern 
\citep{Wang:Zirin:1992,Sobotka:etal:1999}. A converging flow was also found
in MHD simulations of a pore by \citet{Cameron:etal:2007b}. We focus in the 
following discussion on large scale flows in the deeper domain, which 
develop on time scales of several hours to days. 
  
Figure \ref{fig:f2} presents the time evolution of the subsurface flow field 
for the same time steps shown in Figure \ref{fig:f1}. The two columns on
the left show vertical and radial velocity (with respect to the approximate
center of the spot) in a depth of $-3.3$~Mm, which corresponds to the magnetic
field evolution shown in the middle column of Figure \ref{fig:f1}. Here, and
throughout this paper, upflows are represented by positive values and blue
colors, outflows are represented by positive values and red colors. The right
column presents azimuthal averages of the radial flow velocity as function
of distance from the spot center (horizontal axis) and depth (vertical axis).  
Since convective flow velocities vary quite substantially in a $16$~Mm deep
domain from a few $100\,\mbox{m\,s}^{-1}$ near the base to a few 
$\mbox{km\,s}^{-1}$ in the photosphere,
we normalize here the azimuthal flow velocity by the convective rms velocity
at the corresponding depth (averaged over the region outside the sunspot). 
Over time we see the development of a radial mean flow with a weak inflow 
close to the sunspot and a large scale outflow in the region $R>10$~Mm.
Toward the end of the simulation run the azimuthally averaged outflow reaches 
an amplitude of about $50\%$ of the
convective rms velocity, while the inflow reaches values around $15\%$
(faster flows are found in the photosphere).
Furthermore we see a strong trend of declining inflow and increasing outflow
amplitude over time. The presence of the sunspot leads over time to a ring-like
arrangement of convection cells in the periphery of the spot, which was already
found in the more shallow simulation of 
\citet{Rempel:etal:Science,Rempel:2011} and the idealized simulations 
of \citet{Botha:etal:2011}. The result
is a ring of upflowing material with overall reduced vertical flow
velocity and fewer downflow lanes relative to convection further away from
the sunspot (left column). The radial flow patterns (middle and right column)
shows a converging flow in the proximity of the spot and a diverging flow 
further out. At earlier time steps the 
large scale flows near the bottom of the domain are less
pronounced since the re-arrangement of the convection pattern requires
several turn-over time scales and the latter increases substantially with 
depth. For example, the quantity $H_p/v_{\rm rms}$ is about 1 hour in the
middle of the domain, but 6-8 hours near the bottom of the domain.  

Figure \ref{fig:f3} shows for a single time step ($t=46.4$ hours) 
vertical and radial flow velocity for the depth levels of $-11.5$,
$-3.3$ and $-1.3$~Mm. The ring-like arrangement of convection cells is
present at all depth levels shown, the scale of the resulting large scale flow
patterns is increasing with depth as the intrinsic scale of convection is
increasing proportional to the pressure scale height. This tendency is also
manifest in the azimuthal average of the radial flow velocity toward the
end of the simulation run (see Figure \ref{fig:f2} for $t=46.4$ hours). 

\begin{figure*}
  \centering 
  \resizebox{0.95\hsize}{!}{\includegraphics{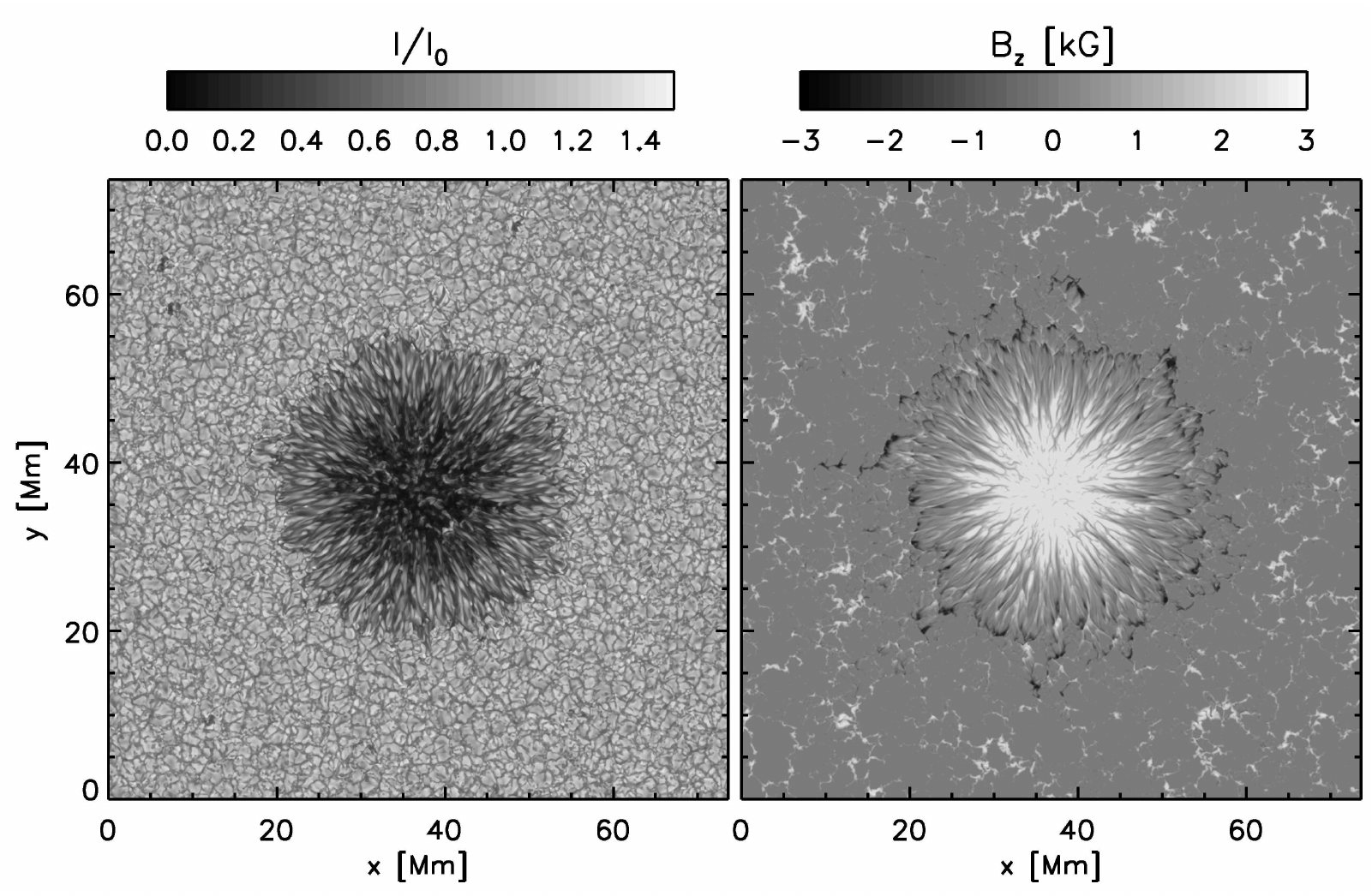}}
  \caption{Intensity image and magnetogram of the sunspot with penumbra for 
    which we present the subsurface flow structure in Figures \ref{fig:f5}, 
    \ref{fig:f6}, and \ref{fig:f8}; long-term averages of the photospheric 
    brightness are presented in Figure \ref{fig:f11}. Displayed are 
    intensity and magnetic field after running the simulation for $24$ hours.
    An animation is available in the online material.\label{fig:f10}
  }
\end{figure*}

To analyze the large scale flows in more detail we focus now
on long-term time averages of the azimuthally averaged  radial and vertical
flow velocity, mass flux and temperature perturbations. To this
end we averaged 19 snapshots between $t=35$  and $t=50$ hours (about 50 
minute cadence). The result is presented in Figure \ref{fig:f4}. Velocities
are normalized by the convective rms velocity, the temperature fluctuation
is defined relative to the mean stratification, both, rms velocity and
reference stratification are computed from the region surrounding the sunspot
near the edges of the computational domain. We indicate the extent of the 
sunspot through field lines of the azimuthally averaged field. The outermost
field line encloses $6.5\cdot 10^{21}$~Mx flux. We contrast this result with 
a sunspot that has an extended penumbra and Evershed flows reaching more 
than $4\,\mbox{km\,s}^{-1}$ on average in Figure \ref{fig:f5} (see
description in Sect. \ref{sect:models} for details). All quantities are the
same as in Figure \ref{fig:f4}, except for the outermost field line which 
encloses $8.8\cdot 10^{21}$~Mx flux. Note that both simulations have 
different domain sizes ($49.152\times 49.152\times 16.384$~Mm$^3$ in Figure 
\ref{fig:f4} and $73.728\times 73.728\times 9.216$~Mm$^3$ in Figure 
\ref{fig:f5}). We present a common
subsection in both figures for better comparison. Figure \ref{fig:f5} presents
a $12$ hour average from $t=12$ to $t=24$ hours, an intensity image at
$t=24$ hours for this simulation is presented in Figure \ref{fig:f10}.

In both cases (Figure \ref{fig:f4}: sunspot without penumbra; Figure \ref{fig:f5}:
sunspot with penumbra) the dominant feature is a deep reaching outflow region 
surrounding the spot with radial mean velocities of about $0.5\,v_{\rm rms}$ 
(panels a). The corresponding large scale upflow (panels b)
coincides with a large scale temperature enhancement of about $0.5\,T_{\rm rms}$
(panels d), indicating a thermal signature typical for a convective flow 
pattern. The amplitude of the radial flow as well as thermal signature is
larger for the sunspot with penumbra. The same is also true for the fraction
of the overturning mass flux found in the azimuthal mean component (panels c) 
which is about $55\%$ ($80\%$) for the sunspot without (with) penumbra in 
$8$~Mm depth. 

The most obvious difference between both sunspot models are present in
photospheric layers along the indicated $\tau=0.1$ level (see also Figure 
\ref{fig:f6} for further detail): in absence of a penumbra we find a 
converging flow for $R<10$~Mm and a diverging flow further out; in presence 
of a penumbra radial outflows are found everywhere in the photosphere.
In absence of a penumbra the converging flow shows a downward extension
in the proximity of the sunspot, which disappears in the presences of a 
penumbra: radial outflows are found at all depth levels. As a consequence
we do not find a situation in which a shallow Evershed flow is stacked on 
top of an inflow cell deeper down as it has been suggested by some
helioseismic inversions \citep{Zhao:etal:2010}.   

Note that in Figure \ref{fig:f4} downflows appears to be located within the 
sunspot, which is caused by the azimuthal averaging: the flow patterns 
presented in Figs. \ref{fig:f2} and \ref{fig:f3} clearly show that downflows
are located at the periphery of the highly fragmented subsurface magnetic 
field structure. The conspicuous temperature enhancement present in 
Figure \ref{fig:f5}d) at the bottom boundary near $R=5$~Mm is a consequence 
of our boundary condition, which keeps the magnetic field fixed within the 
inner most $8$~Mm. 

Radial flows in the upper most $3$~Mm of the domain for both cases are 
presented in Figure \ref{fig:f6}. In contrast to the previous figures we did 
not normalize here the velocity with the convective rms velocity. In the
case of the sunspot without penumbra (panel a) we see outflows at 
photospheric levels for $R>11$~Mm of about $600\,\mbox{m\,s}^{-1}$. 
These flows are a direct 
continuation of the subsurface flow cell identified above and they reach their
peak amplitude in the photosphere in between the $\tau=1$ and $\tau=0.01$ level.
In the case of the sunspot with penumbra (panel b) the upward continuation 
of the deep reaching flow cell coincides with the penumbra and Evershed flow
reaching an amplitude of about $4\,\mbox{km\,s}^{-1}$. In addition we see at 
greater distance
($R>20$~Mm) an additional superficial outflow pattern with amplitudes 
of about to $400\,\mbox{m\,s}^{-1}$ mostly above $\tau=1$. This flow 
cell is neither related to the deep reaching flow component nor a continuation 
of the Evershed flow. 

Figure \ref{fig:f7} shows vertical profiles of
this flow component for both sunspots together with the radial magnetic field
and temperature. The largest flow amplitudes are found mostly above $\tau=1$ and
the velocity peaks in the region with the steepest increase of $B_R$ toward 
the overlying magnetic canopy. Since most of the stronger magnetic field is 
found in the overlying canopy, the convection pattern remains close to 
isotropic. This differs from the situation found within the penumbra, 
where the magnetic field is strong enough to cause substantial anisotropy at 
and below $\tau=1$ 
\citet{Kitiashvili:etal:2009,Rempel:etal:Science,Rempel:2011} and leads to a 
radial flow velocity that peaks near $\tau=1$. In that sense the flow shown
in Figure \ref{fig:f7} is better characterized as overshooting granulation 
that is outward deflected by the overlying inclined magnetic canopy.

In addition strong inflows are present in layers higher
than $\tau=0.001$ regardless of the presence or absence of an penumbra, which
are possibly related to the inverse Evershed flow \citep{Dialetis:etal:1985}. 
This flow component is very robust and present in all numerical sunspot 
simulations we performed to date, however, due to the rather crude treatment 
of physics in those layers a more detailed study of this component would be 
required before stronger conclusions can be drawn 
\citep[see also comments in][]{Rempel:etal:Science,Rempel:2011}.

Overall we identified in our simulations a total of 5 flow components that 
contribute to horizontal flows in the proximity of sunspots. We highlighted
the following flow components in Figure \ref{fig:f6}: (1) The fast
photospheric Evershed outflow in the presence of a penumbra, (2) a converging 
fast photospheric inflow (with a weaker downward extension) in the absence of
a penumbra, (3) a deep reaching outflow cell with average velocities of
about $50\%$ of the convective rms velocity, (4) a superficial outflow
underneath the magnetic canopy, and (5) an inflow above $\tau=0.001$. The
flow components (1) and (2) are obviously strongly affected by the presence
or absence of an penumbra, while the flow components (3) - (5) are of
independent origin.

While the temporal evolution of the large scale flows surrounding the sunspot 
without penumbra (see Figure \ref{fig:f2}, right panels) shows an overall trend
toward increasing outflow velocity, we find the opposite for the sunspot with
penumbra. Figure \ref{fig:f8} presents time averages of the radial flow
velocity from 6-12 and 18-24 hours. The amplitude of flows in the photosphere
is declining by about a factor of two. While in earlier stages the flow 
component (4) showed some connection to the deeper reaching component (3), 
it is mostly disconnected toward the end of our simulation run. We 
return to this aspect in Sect. \ref{sect:dis:evershed_moat}.

\begin{figure*}[t]
  \centering 
  \resizebox{0.95\hsize}{!}{\includegraphics{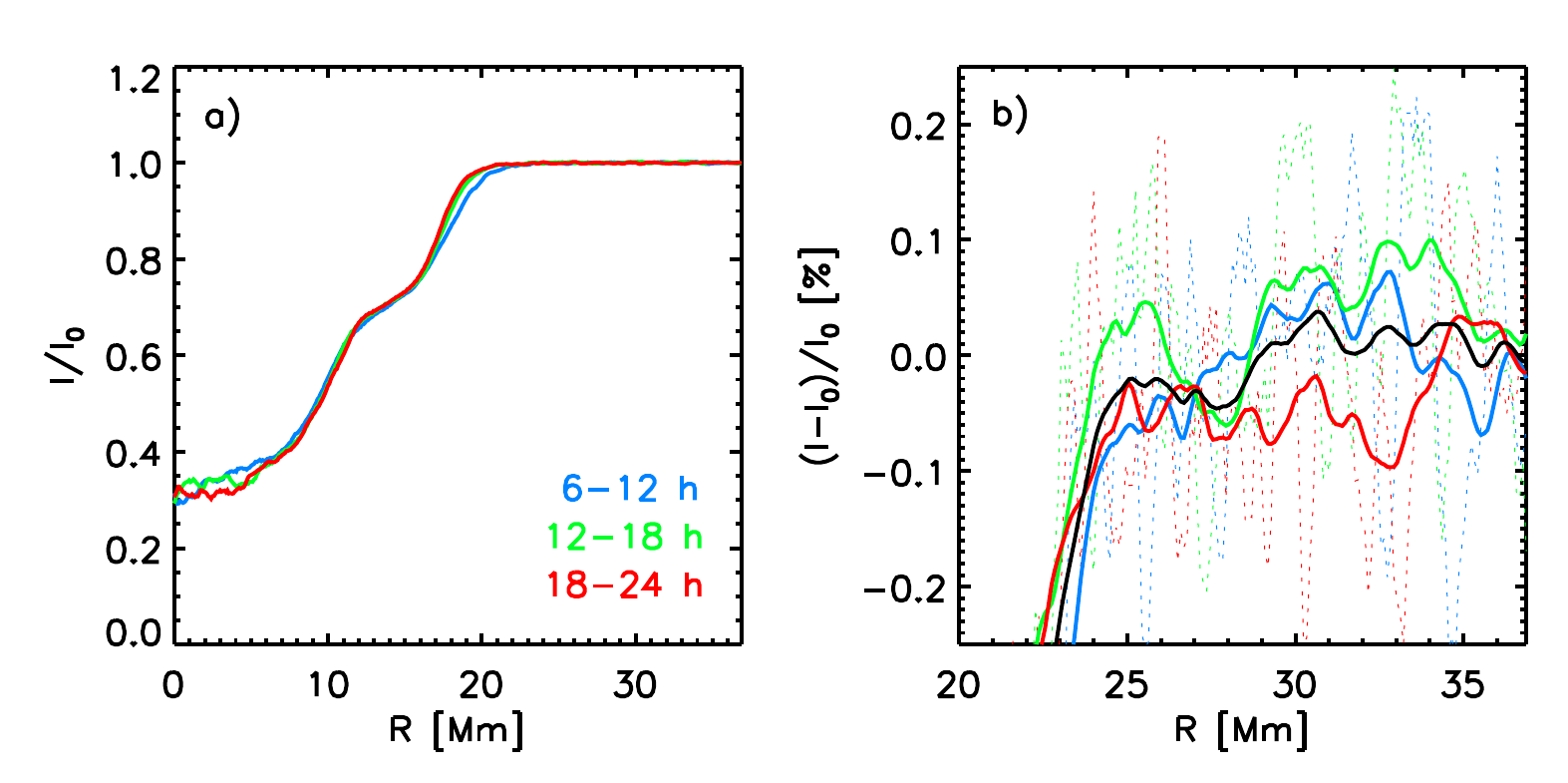}}
  \caption{Temporal averages of the the azimuthally averaged intensity for
    the sunspot presented in Figure \ref{fig:f10}. Panel a) shows the
    Intensity profile from the center of the spot to the edge of the moat 
    region, panel b) presents fluctuation around the mean intensity
    in the moat region. Dotted lines indicate the fluctuations present in 6 hour
    azimuthal averages. Solid lines are in addition averaged with a $2.5$~Mm
    window in the radial direction to highlight potential trends of intensity
    with radius. The black line shows the average from $6-24$~h.\label{fig:f11}
  }
\end{figure*} 

\subsection{Origin of large scale subsurface flows}
\label{sect:flow-orig}
The large degree of similarity between Figs. \ref{fig:f4} and \ref{fig:f5}
strongly suggests an origin of the deep reaching flow component (3) that 
is independent from the Evershed flow. It is however non-trivial to clearly 
disentangle 
in the numerical simulations presented here the different processes that can
give rise to large scale flows. There are primarily two effects: 1.) The sunspot
imposes a geometric constraint on the surrounding convection pattern, which
forces adjacent convection cells to align in a ring like pattern resulting
in a mean flow. 2.) Reductions in the surface brightness lead to less downflows
in the proximity of the spot. Since downflows provide low entropy material
the result is an increase of the mean temperature driving a large scale flow
pattern through buoyancy. We see evidence for both processes at work. Figs.
\ref{fig:f2} and \ref{fig:f3} show clearly the ring like arrangement of 
convection cells and also the strong reduction of downflow lanes within
this region. To get an independent estimate of the strength of both 
processes we report here on a series of idealized simulations in a $49.152$~Mm 
wide and $8.192$~Mm deep domain. 

In the first setup we cut out a central cylinder with a radius of $8$~Mm to 
independently investigate the geometric effect. To this end
we artificially set all velocities to zero within this region and replace
the stratification with the mean stratification found outside this cylinder.
In the second setup we set the radiative cooling function to zero within a
central cylinder of $8$~Mm to investigate flows resulting from the blockage of 
heat flux. A combination of geometrical and thermal effects can be realized
by considering a cone instead of a cylinder.

The resulting azimuthally averaged flows are presented in Figure 
\ref{fig:f9}. In the case of the pure geometric constraint (panel a) 
we see the development of a large scale flow cell, with flows 
converging toward the central cylinder for $R<12$~Mm and diverging flows 
further out. Overall the extent of the region with mean flows is about $20$~Mm,
which is about $2.5$ times the radius of the central cylinder. However, the 
amplitude of the outflow is with $10-15\%$ of the convective rms velocity 
rather weak, larger values are found in the inflow cell in close proximity 
of the central cylinder. 
Heat flux blockage (panel b) leads to outflows at all depth levels in
the proximity of central cylinder, even though further out we see the
development of an inward directed flow cell. The mean flow speed is comparable
to convective velocity in the upper most $2$~Mm but drops to values of less
than $50\%$ of the convective rms velocity in deeper layers. In this experiment
we assumed complete blockage of the heat flux, which clearly overestimates
the influence of a sunspot. A combination of both effects can be realized 
by placing an upward opening cone as obstacle into the center of the domain
(bottom panels). The mean flow velocity increases with opening angle and
values of about $50\%$ of the convective rms velocity can be achieved with
opening angle of $30\deg$ (panel d). From these idealized experiments 
we can estimate that the bulk of the outflows results from blockage of heat 
flux, while geometric effects can have a contribution of up to $30\%$. 

The results from the idealized experiments show a good qualitative agreement 
in terms of amplitude and outflow pattern with the flows presented in Figs. 
\ref{fig:f4} and \ref{fig:f5} (primarily flow component 3). In the
case of the sunspot with penumbra (Figure \ref{fig:f5}) the region with 
enhanced subsurface temperature is as expected centered underneath the 
penumbra with reduced heat loss. Interestingly the region with enhanced 
temperature in (Figure \ref{fig:f4}) is found in the same location, despite 
the absence of a penumbra. Furthermore the subsurface flow cells have in both 
cases about the same radial extent despite the fact that the sunspot with 
penumbra has almost twice the diameter. This is likely a consequence of
the converging flow cell surrounding the sunspot without penumbra, which is
shifting the location of diverging flow pattern further outward relative to the
spot boundary. The converging flow cell is not present in the sunspot with 
penumbra.

\begin{figure*}
  \centering 
  \resizebox{0.9\hsize}{!}{\includegraphics{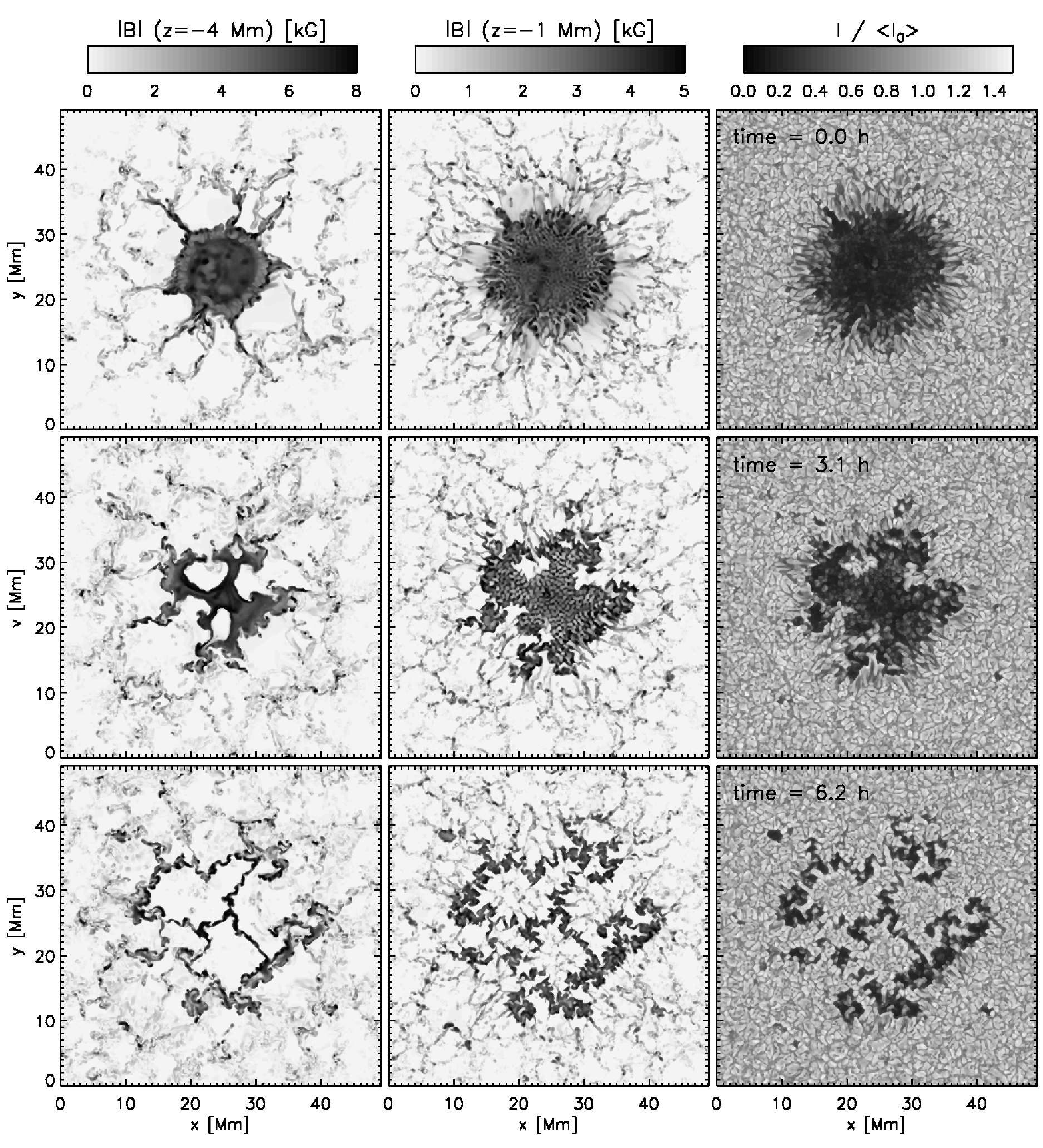}}
  \caption{Temporal evolution of a sunspot in a $6.144$~Mm deep domain.
    The bottom boundary is symmetric in all three mass flux components.
    The sunspot is decaying on a time
    scale of about $3-6$ hours due to interchange instabilities developing
    near the bottom of the domain. The field free plasma intruding near the
    bottom boundary leads to buoyant upflows that develop islands of 
    granulation withing the umbra of the spot. After about $6$ hours
    most of the magnetic field is dispersed into structure of the size of a 
    few pores.\label{fig:f12}
  }
\end{figure*}

\begin{figure*}
  \centering 
  \resizebox{0.9\hsize}{!}{\includegraphics{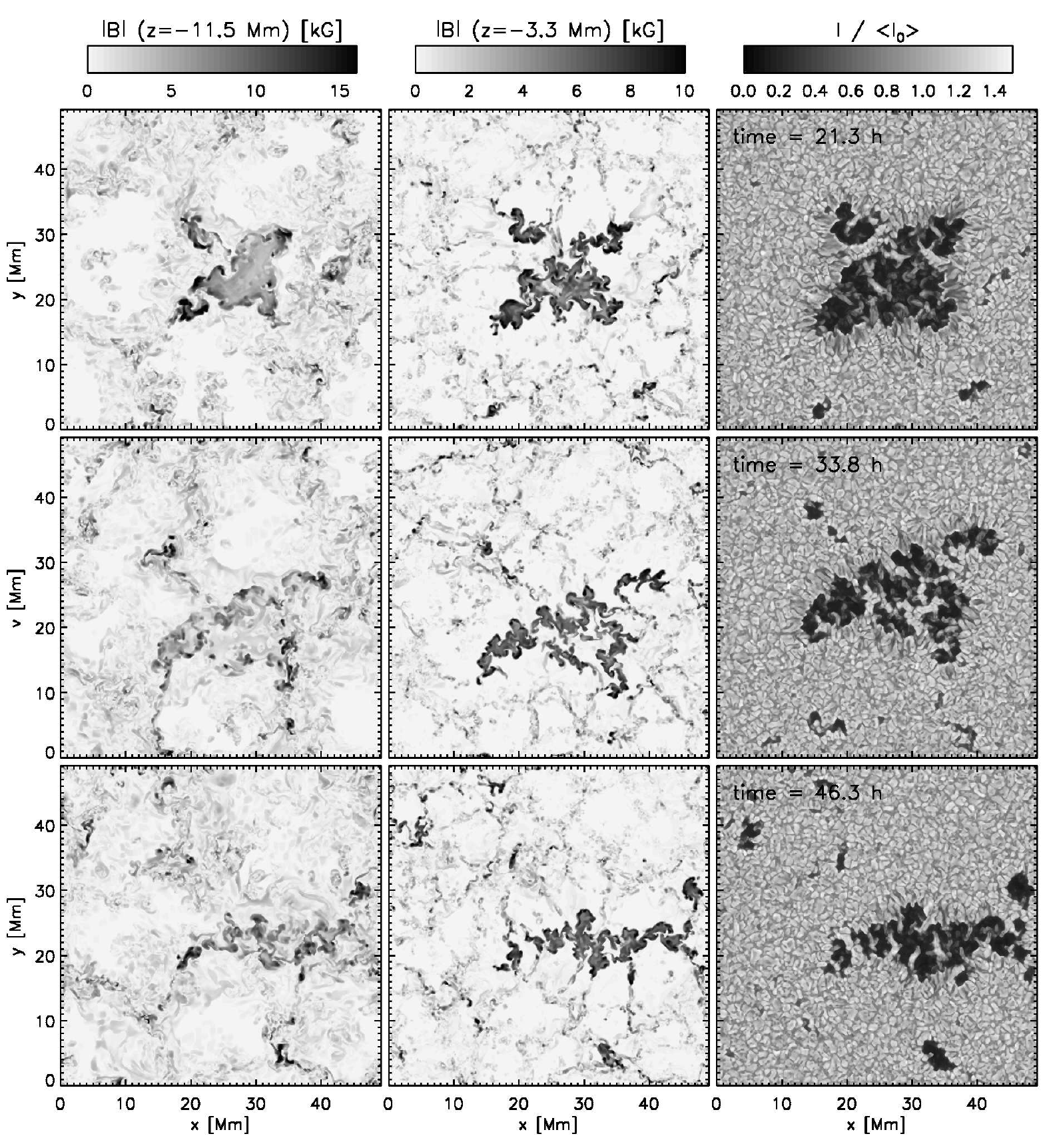}}
  \caption{Temporal evolution of a sunspot in a $16.384$~Mm deep domain using
    an open bottom boundary condition (symmetric in all three mass flux 
    components). The simulation was restarted from the snapshot at $t=8.8$~h 
    shown in Figure \ref{fig:f1}, the elapsed time is relative to $t=0$~h to be
    comparable to Figure \ref{fig:f1}. The open boundary 
    condition leads to a stronger deformation and fragmentation, compared to
    the reference run shown in Figure \ref{fig:f1}, but even at $t=46.3$ hours 
    the larger fraction of the flux remains concentrated within the main 
    sunspot (see Figure \ref{fig:f14}). Compared to
    the $6$~Mm deep domain presented in Figure \ref{fig:f12}, dynamical time 
    scales near the bottom are about $6-8$ times longer. An animation is 
    available in the online material.\label{fig:f13}
  }
\end{figure*}

\begin{figure*}
  \centering 
  \resizebox{0.95\hsize}{!}{\includegraphics{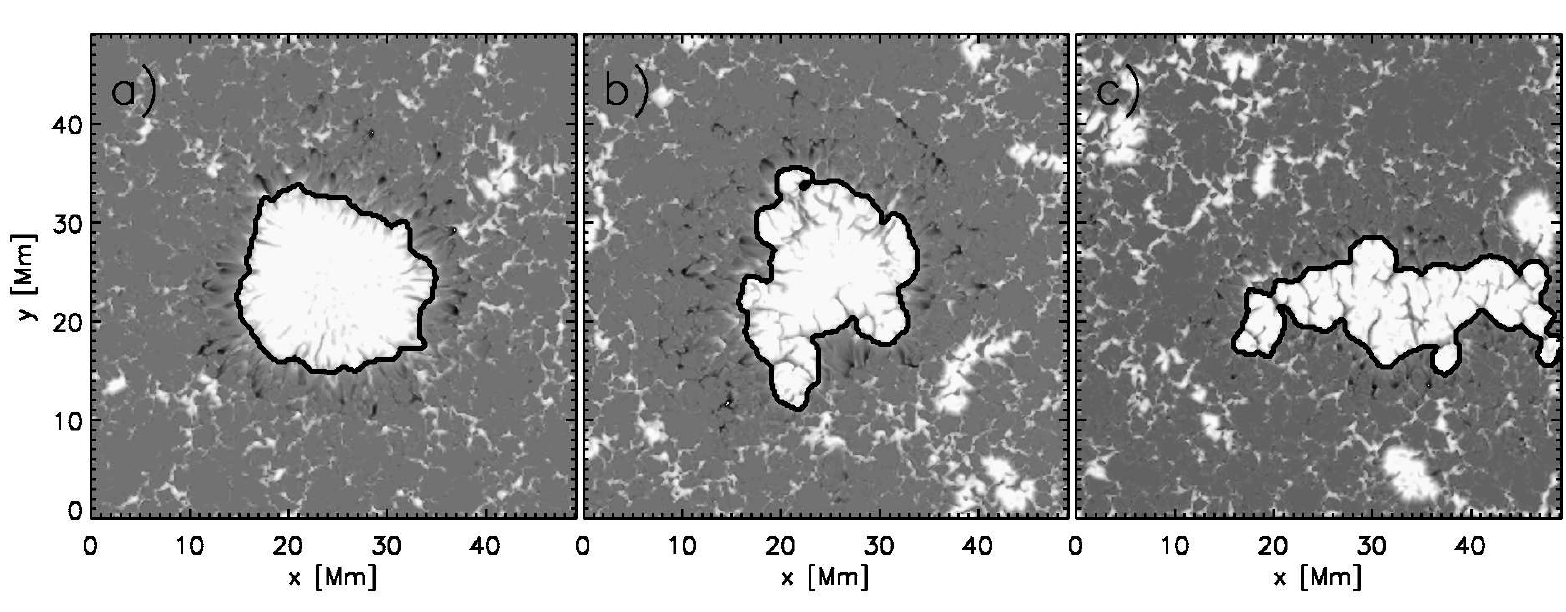}}
  \caption{Influence of bottom boundary condition on sunspot decay. Panel
    a) shows the initial state at $t=8.8$~h from which we started the
    simulation run with open boundary condition, panel b) shows the result
    for the closed boundary condition at $t=46.4$~h, and panel c) for the
    open boundary condition at $t=46.3$~h. Presented is a magnetogram at
    $\tau=1$ together with a mask we used to compute the flux and area of 
    the spot. The values are a) $9.1\cdot 10^{21}$~Mx, b) $7.8\cdot 10^{21}$~Mx,
    and c) $6.5\cdot 10^{21}$~Mx for flux and a) $275$~Mm$^2$, b) $257$~Mm$^2$,
    and c) $236$~Mm$^2$ for the area.\label{fig:f14}
  }
\end{figure*}

\subsection{Are there bright rings around sunspots?}
Whether sunspots are surrounded by bright rings that account for some of
the energy flux blocked by the lower brightness in umbra and penumbra has
been extensively discussed in the literature. We refer here to 
\citet{Rast:etal:2001} and references therein for a detailed discussion of both 
theoretical and observational aspects of this problem. On the theoretical side
it has been argued that the thermal conductivity and heat capacity of the 
convection zone are sufficiently large to absorb temporary thermal 
perturbations caused by sunspots without becoming visible as brightness
change in the photosphere \citep{Spruit:1982a,Spruit:1982b,Foukal:etal:1983,
Chiang:Foukal:1985,Spruit:2000:SSRv}. On the
observational side the main challenge lies in the proper separation of this
effect from Facular brightening, which is a pure surface effect. 
\citet{Fowler:etal:1983} found bright rings in the $0.1-0.3\%$ range, the
more recent study by \citet{Rast:etal:2001} brightness enhancements of
$0.5-1\%$, which are attributed to convective origin.

We analyze here the numerical simulation in the 
$73.728\times 74.728\times 9.216$~Mm$^3$ 
domain, which has a more realistic penumbra and a moat
region of at least $20$ Mm surrounding the spot. A snapshot of the intensity 
for a vertical ray is presented in Figure \ref{fig:f10}, 
Figure \ref{fig:f11} shows azimuthally averaged intensity profiles 
for this sunspot model. We show three consecutive 6 hours 
temporal averages of the bolometric intensity for a vertical ray. Panel a)
shows intensity profiles from the umbra into the moat region, panel b)
fluctuations within the moat region on a different scale. We do not see 
evidence for a systematic variation of intensity with radius in the moat 
region. For the individual 6 hour averages systematic trends are smaller than
$0.1\%$ and these trends are different for consecutive 6 hour averages. The
dark solid line shows the average of the intensity from $6-24$ hours. Here
we see a weak ($0.05\%$) enhancement of intensity in the moat region with a 
peak at about $R=30$~Mm, however, given the substantial variation between the
6 hours averages it would require a longer simulation run to determine
the robustness of this trend. The presented intensity profiles are based
on azimuthal averages that take into account all magnetized areas in the
moat region. Since our numerical resolution of $48$~km in the horizontal 
direction in combination with grey radiative transfer is not sufficient
for resolving Faculae, we do not have to correct for Facular
brightening. Nevertheless, we repeated the analysis after cutting out 
magnetized regions. If we consider only regions with $\vert B_z\vert<500$~G
the results remain essentially the same. If we lower the threshold to $250$  G
we see a bright ring with about $0.3\%$ amplitude. Since most vertical 
magnetic field is concentrated in the darker downflow lanes and more unsigned
vertical flux is present in the proximity of the sunspot the latter is an
artifact of removing more dark downflow lanes from the average in the inner
moat region compared to the outer. 

We also note that the total energy flux in our domain is reduced compared to
quiet sun convection, since the convective energy flux crossing the bottom
boundary is free to adjust (we specify the entropy in inflow regions, but 
the mass 
flux crossing the boundary is not constrained). In addition most of the 
energy flux that is blocked by the umbra of the sunspot does not enter the 
computational domain in the first place due to our choice of the bottom 
boundary condition. Nevertheless, some of the energy flux entering the domain
underneath the penumbra has to be diverted away from the spot, which is, as we
discussed above, the primary reason for the presence of the large scale 
subsurface flow cell surrounding the sunspot.
The latter process is associated in the deeper layers with a temperature 
excess reaching about $50\%$ of the convective rms temperature fluctuation
(see Figure \ref{fig:f5}d).
The absence of bright rings at photospheric levels clearly shows that this
temperature perturbation remains well hidden beneath the photosphere.

\section{Influence from bottom boundary condition}
\label{sect:bottom-bnd}
The simulation we reported on so far in Sect. \ref{sect:results} as well as
previous simulations by \citet{Rempel:etal:2009} and 
\citet{Rempel:etal:Science} were based on a bottom boundary that sets all
velocities to zero in regions of strong field, but allows for convective
motions crossing the boundary in regions with weak field. In the latter,
velocity components are symmetric across the boundary in downflow regions,
while the velocity is vertical (horizontal components antisymmetric) in
upflow regions. In shallow domains with about $6$~Mm depth this boundary 
condition is a necessity 
to prevent a rather quick decay of a sunspot within a few hours. We demonstrate
this here by a numerical experiment performed in a $6.144$~Mm deep domain using
an open bottom boundary, which imposes a symmetric boundary for all
three mass flux components. In regions with $\vert B_z\vert<2.5$~kG we impose
a constant total pressure at the bottom boundary, in regions with
$\vert B_z\vert>2.5$~kG we extrapolate the pressure stratification. The latter
is necessary in a shallow domain to prevent long lasting inflows that would 
destroy the sunspot on even shorter time scales than reported here.

Figure \ref{fig:f12} shows the evolution of the sunspot in the $6.144$~Mm deep 
domain. We restarted this simulation from a run which was previously evolved
with a closed boundary condition in regions with $\vert B_z\vert > 2500$~G,
leading to a nearly axisymmetric stable sunspot (top panels of Figure 
\ref{fig:f12}). After only a few hours we observe a decay of the spot that is
primarily driven through interchange instabilities. After $3$ hours we see 
the clear evidence of intrusions of field free plasma near the bottom 
boundary, which pushes buoyantly upward within the magnetized region and
becomes visible in the photosphere through granulation patches showing up
at the periphery of the umbra. After $6$ hours these regions have expanded 
to patches of $5-10$~Mm diameter and most of the umbra is dispersed
along the periphery of the patches.

In Figure \ref{fig:f13} we contrast this result with a repetition of the same
experiment in a $16.384$~Mm deep domain. To this
end we start from the snapshot at $t=8.8$ hours shown in Figure \ref{fig:f1}
and evolve the simulation for $37.5$ hours with the open boundary 
condition. In contrast to the experiment performed in the $6.144$~Mm deep domain
we impose a constant total pressure everywhere in the domain.
Figure \ref{fig:f13} presents 3 snapshots about $12.5$ hours
apart from each other, the quantities shown are the same as in Figure 
\ref{fig:f1}. Suddenly opening the bottom boundary beneath the sunspot leads
to an upflow strongly weakening the magnetic field, however, due to the
deeper domain this upflow does not continue into the photosphere. After the
initial weakening in the deep domain the trunk of the sunspot
fragments into smaller flux bundles. The photospheric signature of this
process is an increased rate of deformation and fragmentation of the sunspot
compared to the reference in Figure \ref{fig:f1} with the closed boundary,
but even at $t=46.3$ hours ($37.5$ hours of evolution with open boundary 
condition) most of the flux remains concentrated together. 
The overall slower decay rate in the deeper domain compared to the 6 Mm deep 
domain discussed above is consistent with the increase of the convective time 
scale by a factor of about $6-8$ compared to the shallow domain. In Figure 
\ref{fig:f14} we compare the flux remaining in the sunspots toward the
end of our simulation runs. To evaluate the flux of the sunspots we compute
the flux within the shown $1.25$~kG contour line, which is based on a 
smoothed magnetogram (Gaussian with FWHM of $960$~km). 
In the case of the closed (open) boundary condition
about $86\%$ ($71\%$) of the initial flux at $t=8.8$ hours (shown in panel a) 
is still concentrated together. At the same time the average field strength is
dropping from $3300$~G to $3040$~G ($2760$~G), so that the actually remaining 
spot area is $93\%$ ($85\%$). Overall the decay rate is about a factor of 2 
larger for the open boundary condition. Averaged over the $37.5$ hours of the 
simulated decay the corresponding flux loss rates are 
$0.8\cdot 10^{21}$~Mx~day$^{-1}$ ($1.5\cdot 10^{21}$~Mx~day$^{-1}$).
These values are about a factor of $10$ larger than those observed by
\citet{Martinez-Pillet:2002}, who found values around
$0.6-1.44\cdot 10^{20}$~Mx~day$^{-1}$. For a substantially larger sunspot
\citet{Kubo:etal:2007} found values of $0.6-0.8\cdot 10^{21}$~Mx~day$^{-1}$.
A detailed comparison to observations is currently only of limited practicality
since our simulated sunspots shown in Figure \ref{fig:f1} and \ref{fig:f13} are 
essentially penumbra free ``naked'' sunspots and most reported sunspot decay
rates refer to spots with fully developed penumbrae. However, a direct
comparison between both simulations gives important clues about the influence 
of the bottom boundary condition.

Since we start our simulations from
a thermally relaxed HD simulation with magnetic field added to it, initially
magnetic field and surrounding flow patterns are not necessarily dynamically
consistent. In the case with closed boundary conditions (Figure \ref{fig:f1})
the flow field is forced to adjust to the presence of the field, while in
the case of open boundary conditions the magnetic field adjusts to the
flow field, explaining the substantial change in overall shape of the sunspot
in Figure \ref{fig:f13}. As a consequence the experiment shown in 
Figure \ref{fig:f13} likely overestimates the decay rate due to this initial
adjustment state. 

Overall our set of experiments demonstrates that the constraints imposed by 
the bottom boundary condition on the time evolution of sunspots can be relaxed 
by performing simulations in deeper (more computationally expensive) domains.
A $16$~Mm deep domain is consistent with sunspot life times of a few days,
which is still much shorter than the life time of most observed sunspots.
Extrapolating this result a life time of a week or more should be reached in the
$30-50$~Mm depth range. The overall decay rate is likely also strongly dependent
on the initial state, which can be obtained only in a self-consistent manner
through a flux emergence simulation describing the sunspot formation process. 

\section{Discussion}
\label{sect:discussion}
\subsection{Flow components around sunspots}
\label{sect:dis:flow}
We find large scale outflows around sunspots as a robust result from
a series of numerical simulations including idealized experiments. We 
identified 5 different flow components:
\begin{enumerate}
  \item The Evershed flow closely related to penumbral fine structure and 
    driven by magneto convection in an inclined magnetic field. 
  \item In the absence of a penumbra a converging flow caused by enhanced 
    radiative cooling of granules adjacent to the umbra (flow extends with
    reduced amplitude to deeper layers).
  \item A deep reaching large scale flow that originates from
    a combination of geometric constraints and blockage of heat flux, with the
    latter providing the dominant contribution. The underlying convective
    flow morphology in the periphery of the sunspot is a ring-like arrangement 
    of convection cells, which leads to large scale outflows reaching about
    $50\%$ of the convective rms velocity over the depth range currently 
    accessible through numerical simulations (down to $16$~Mm depth).
  \item A shallow outflow component found in the plage region surrounding
    the sunspots. This flow is mostly limited to the photosphere and 
    is best characterized as overshooting granulation that is outward 
    deflected by the overlying inclined magnetic canopy.
  \item An inflow in levels higher than $\tau=0.001$ located near the outer 
    edge of the umbra or above the penumbra (if present).
\end{enumerate}
By comparing numerical simulations of sunspots with and without penumbra 
we identified an origin of the flow components (3) to (5) that is independent
from the existence of a penumbra and Evershed flow. Through a series of
idealized experiments we demonstrated that the flow component (3) arises
from a combination of geometric constraints and blockage of heat flux, with
the latter playing the dominant role.

In models without 
penumbra we find a two cell flow pattern around the sunspot, consisting
of a converging flow in the proximity of the spot (2) and a diverging flow 
further out (3), which merges in the photosphere with the flow pattern (4).
The dominant flow component in this case is (3). 

In models with penumbra the flow pattern (2) is not present and replaced with
(1). Flow pattern (1) merges continuously in the deeper layers with (3),
resulting in outflows at all depth levels underneath the penumbra. Flow
pattern (4) is in this case somewhat detached and leads to a superficial
flow component outside the sunspot. The amplitude of this flow component
is declining throughout the simulation from about $600$~ms$^{-1}$ to less 
than $300$~ms$^{-1}$. 

The inflow component (5) in higher layers is present in both cases and possibly
related to the inverse Evershed flow \citet{Dialetis:etal:1985}. This flow
is very robust and present in all numerical sunspot simulations we performed
to date, however, due to the rather crude treatment of physics in those layers
a more detailed study of this component would be required before stronger
conclusions can be drawn 
\citep[see also comments in][]{Rempel:etal:Science,Rempel:2011}.

The Evershed flow component (1) has been discussed with great detail in
previous publications about MHD simulations
\citep{Heinemann:etal:2007,Scharmer:etal:2008,Rempel:etal:2009,
Kitiashvili:etal:2009,Rempel:etal:Science,Rempel:2011} and is
well studied in observations 
\citep[see e.g.][]{Solanki:2003,Thomas:Weiss:2004}, 
we focus the following discussion primarily on the flow components (2) to (4).

The inflow component (2) has been observed in pores as well as spots without
penumbra \citep{Wang:Zirin:1992,Sobotka:etal:1999,VDominguez:etal:2010} and
has been identified in 3D MHD simulations of pores \citep{Cameron:etal:2007b},
where it plays a crucial role in maintaining the magnetic structure as well 
as forming it \citep{Kitiashvili:etal:2010:pore}.

The dominant subsurface outflow component (3) is essentially a convective flow
that is modified due to a combination of geometric constraints together
with heat flux blockage caused by the presence of the sunspot. Associated
with this flow is a ring of temperature excess around the sunspot, which is 
similar to the flow pattern and related thermal perturbations that were 
suggested by \citet{Meyer:etal:1974}, although we see this flow on scales 
smaller than supergranulation. Despite a thermal signature in deeper 
layers with amplitudes comparable to those of convective rms temperature 
fluctuations we do not see compelling evidence for a bright ring surrounding 
the sunspot in the photosphere. 

Apart from being a shallow photospheric moat flow component, the flow 
pattern (4) could be possibly also related to observed flows along the
magnetic canopy \citep[see e.g.][and references therein]{Rezaei:etal:2006}.
Since we see this effect in our simulations regardless of the presence or 
absence of a penumbra this component is not an extension of the Evershed 
flow in our simulations.

The flow pattern we find around the sunspot without penumbra (converging
flow in close proximity, diverging flow further out) is similar to
flows found in axisymmetric sunspot simulations such as 
\citet{Hurlburt:Rucklidge:2000}, \citet{Botha:etal:2006}, and 
\citet{Botha:etal:2008}. It has been speculated that in the
case of a sunspot with penumbra only a shallow outflow (Evershed flow) is 
added on top of this flow pattern, while a converging flow remains in
deeper layers \citet{Zhao:etal:2010}. We do not find any evidence for such a
layered flow structure in our simulations. In the case of a sunspot
with penumbra the inflow cell adjacent to the spot disappears and the spot 
is surrounded by outflows at all depth levels (a combination of the flow 
patterns (1) and (3)). This is consistent with the underlying driving 
mechanisms of these flows: In the absence of a penumbra granules in the 
proximity of the umbra have enhanced radiative losses, creating cool downflows
that extend to the bottom of the domain and maintain in the proximity of the
sunspot a converging convective flow pattern. In the presence of a penumbra 
the region with enhanced radiative loss is replaced by a region with reduced
radiative loss, which leads to an overall reduction of cool downflows. In
addition the fast near surface Evershed flow leads to a preferred draining
of these downflows toward the outer edge of the penumbra. The consequence is
a temperature enhancement in the subsurface layers driving a broad upflow 
underneath the penumbra that results in outflows at all depth levels.
This flow patterns is consistent with a recent inversion by 
\citet{Gizon:etal:2009,Gizon:etal:2010:err} that reports on outflows in the 
upper most $4.5$~Mm. A recent result by \citet{Featherstone:etal:2011:JPC}
indicates two contributions to the large scale flows surrounding sunspots: a 
superficial near photospheric and a deeper reaching component that peaks at 
about 5 Mm depth and essentially disappears in depth greater than $9$~Mm.

Our sunspot models are set up in way that they yield almost quasi-stationary
solutions, which implies that they are likely best compared to very
stable, almost circular sunspots with little evolution over time scales of 
a few days. It is possible that the subsurface flow structure is much more 
complicated around more complex active regions with rapid evolution.

\subsection{Connection to observed moat flows}
\label{sect:dis:evershed_moat}
In observations moat flows are typically defined as outflows that are present 
from the outer edge of a sunspot (penumbra) to about 2 sunspot radii. Typical
flow velocities are around $500$~ms$^{-1}$ \citep{Brickhouse:Labonte:1988,
Sobotka:Roudier:2007,Balthasar:Muglach:2010}.
Over the past years it has been discussed in the literature whether Evershed 
and moat flow are connected or of independent origin. 
\citet{SDalda:MPillet:2005} and \citet{Cabrera-Solana:etal:2006} found
examples of moving magnetic features in the moat region that show a connection
to penumbral filaments. \citet{VDominguez:etal:2008} studied several examples
of complex active regions and found a very close connection between 
the presence of penumbrae and moat flows. Furthermore, moat flows were only
found near penumbrae perpendicular to the sunspot border and absent near 
tangential penumbrae. Moat flows were also absent when there was a polarity
inversion line in the proximity of the sunspot. \citet{Zuccarello:etal:2009}
reported on moving magnetic features diverging from a sunspot without
penumbra, indicating an origin of horizontal outflows independent from a 
penumbra and Evershed flow. 

In the simulations presented here we find sunspots surrounded
by outflows in the photosphere with velocities in the $300-600$~ms$^{-1}$
range that extend to about 2 sunspot radii, regardless of the presence
or absence of a penumbra and the associated Evershed flow.

In the absence of a penumbra we find in close proximity of the spot a 
ring of inflowing plasma, which has been also observed around pores by 
\citet{Wang:Zirin:1992,Sobotka:etal:1999,VDominguez:etal:2010}. 
In addition \citet{Sobotka:etal:1999,VDominguez:etal:2010} observed a 
diverging flow further away that is related to a ring-like arrangement
of ``centers of positive divergence'' around the pore, which is similar to 
the preferred ring-like arrangement of convection cells we find around our 
simulated sunspots over a depth range of more than $10$~Mm. In that sense
observations likely see the ``tip of the iceberg'' of this flow structure.  
Large scale outflows are also observed around ``naked'' (penumbra free)
sunspots \citep{Zuccarello:etal:2009}, which we would expect as this process is
largely scale invariant. 

Interestingly, our sunspot with penumbra has an
on average a weaker outflow at photospheric levels compared to the
penumbra free sunspot. In addition the outflow results primarily from the
very superficial flow component (4) and we observe a trend of declining
flow speed throughout the simulation, which is opposite to trend we find
in the simulation with the penumbra free sunspot. At later stages 
($>12$ hours) of the simulation with penumbra we find a 
rather sharp decline of the Evershed flow and subsurface flow pattern (3) at 
about $R=22$~Mm, which results in a downflow that collects most of the 
horizontal mass flux present. This downflow is in part driven by cool material 
deposited there by the Evershed flow (cool plasma forming in the penumbra is 
transported preferentially outward), and in part a consequence of opposite 
polarity magnetic flux accumulating near the outer edge of the penumbra that 
diverts horizontal flows downward. The lower temperature is
evident from Figure \ref{fig:f5}(d), the opposite polarity flux causing a 
polarity inversion line around the sunspot from Figure \ref{fig:f10}. The
absence of moat flows in a region with a polarity inversion near the outer
edge of the penumbra was also observed by \citet{VDominguez:etal:2008}. If this
shielding effect would be weaker it is conceivable that the Evershed flow
component (1) and deep flow component (3) have also larger contributions 
further out, leading to a deeper reaching and faster moat 
flow around the sunspot with penumbra as it is partially indicated during
earlier stages of this simulation (see Figure \ref{fig:f8}a). However,
it would require additional numerical experiments to clearly quantify the 
role of a magnetic inversion line in our simulations.

While we find large scale outflows around sunspots as a robust result
regardless of the presence or absence of a penumbra, it is also possible that
we are still missing some processes contributing to larger scale flows, in
particular on scales of supergranulation that are still not well captured
by the extent of our simulation domain.
Independent from that the simulations presented here clearly show that there
are likely different contributions to large scale flows like the moat flow,
and some of these contributions are independent from the Evershed flow
(components 3 and 4). Based on our findings we conjecture that there is 
likely no clear ``yes'' or ``no'' answer to the question of whether moat and 
Evershed flow are related. To some extent this connection is also subject to 
interpretation and the adopted definition of ``moat flow''. For example 
\citet{VDominguez:etal:2010} did not refer to the outflow they observed around
pores as moat flow, while we see strong indications that those flows have 
likely a much deeper reaching structure and are in part related to blockage 
and modification of convective energy transport around sunspots. 

Due to the substantial differences in the depth extent of the individual 
outflow components we identified (components (1) and (4) are superficial,
while (3) is deep reaching), a clarification of the Evershed-moat flow
relationship has to consider the deeper reaching subsurface flow 
structure through helioseismology \citep[see, for example, recent results by][]
{Gizon:etal:2009,Gizon:etal:2010:err,Featherstone:etal:2011:JPC}. 

\subsection{Robustness of flow structure}
\label{sect:dis:robustness}
The large scale flows we discussed in previous sections reach scales comparable
to the computational domain size. It is therefore essential to discuss to
which degree the flow structure could be influenced by the overall simulation
setup including boundary conditions. A strong dependence on boundary conditions
has been found for example in the 2D axisymmetric simulations of 
\citet{Botha:etal:2008}. We have shown already that the magnetic boundary 
conditions strongly influence the properties of the magnetic sunspot structure.
While the bottom boundary primarily affects the long-term stability and
evolution of the sunspot, the top boundary condition sets the overall extent of
the sunspot penumbra including the associated Evershed flow. With regard to 
large scale flows the key question is whether these flows are a response
to the magnetic structure present in the simulation and are driven by resolved
physical processes within the computational domain or whether they are in 
addition substantially influenced by the hydrodynamical boundary conditions.

Our top boundary condition is located about 700~km above the quiet sun $\tau=1$
level and therefore a density contrast of about three orders of 
magnitude away from the photosphere. Although flow velocities can be 
substantial (larger than 10 km~s$^{-1}$) at this boundary, the associated mass 
and momentum fluxes are negligible and no substantial feedback on flows
in the photosphere and below takes place. The only flow component potentially
affected by this boundary condition is the component (5). Both, the Evershed 
flow (1) and inflow in the absence of penumbra (2) are driven by resolved 
physical processes within the computational domain. This is also the case for
the flow component (4), which is in addition less dependent on the magnetic
boundary condition than components (1) and (2).

The largest scale flow we find in our simulations is the deep reaching flow
component (3). In all of the simulations discussed this flow extends to the
bottom boundary and at least in the domains that are only 50 Mm wide the
radial extent of this flow becomes comparable to the horizontal domain size.
The potential influence of the horizontal and bottom boundary conditions on this
flow could be investigated by either changing the boundary conditions or by 
comparing simulations of similar sunspots in differently sized domains. We have
done here the latter. Figures \ref{fig:f4} and  \ref{fig:f5} compare flow
systems in 2 simulations with quite different domain size, only a common
subsection is shown. In Figure \ref{fig:f4} the simulation domain extends to
about 15.5 Mm depth, i.e. the mass in the part not shown exceeds the
mass in the part shown by about a factor of 10! Nevertheless the flow structure
of component (3) is very similar in both cases, and in particular the absence
of a converging return flow near the bottom boundary is a robust result
(such a flow was only present as a transient during early evolution stages
shown in Figure \ref{fig:f2}). 
Similarly the horizontal extent of the domain does not influence the flow
structure substantially. Even though the domain extends in Figure \ref{fig:f5} 
to a radial position of 37~Mm, the radial extent of the large scale flow is
similar to that found in Figure \ref{fig:f4}. We have done several additional
simulations we did not report here, in particular a simulation of the sunspot
shown in Figure \ref{fig:f4} in an only 8~Mm deep domain and a series of
simulations with different grid resolution of the sunspot shown in Figure 
\ref{fig:f5} in a domain only 49.152~Mm wide and 6.144~Mm deep. In all cases
we find with regard to the axisymmetric flow components similar results
when we compare the overlapping parts of the simulation domains. Additional 
evidence that the horizontal boundary conditions do not matter too much for 
this flow component comes from the ``double sunspot'' simulation of 
\citet{Rempel:etal:Science,Rempel:2011}. Due to the horizontal periodicity
this simulation assumes sunspots of alternating polarity in the x and same
polarity in the y direction. While this strongly influences the structure
of the penumbra and Evershed flow, the subsurface flow pattern corresponding
to flow component (3) remains mostly unaffected 
\citep[][see Figure 20]{Rempel:2011}.

A related concern is the temporal evolution of large scale flows as our
simulations address at this point mostly rather short lived sunspots. This is
not a major concern for the flow components (1), (2) and (5) which are 
established within a few hours of simulation time and did not show any
substantial variation over the time frame of 1-2 days we covered. The flow
components (3) and (4) show a temporal evolution which is presented in
Figures \ref{fig:f2} and \ref{fig:f8}. The rather strong variation of component
(4) shown in Figure \ref{fig:f8} is limited to this particular sunspot and
likely related to the magnetic field evolution in the moat region as discussed
in Section \ref{sect:dis:evershed_moat}. We did not observe a substantial 
variation 
of this flow component around the sunspot without penumbra. Since the flow
component (3) essentially extends from the photosphere to the bottom boundary
of our domain it covers regions with a substantial variation of the intrinsic
convective time scales. The quantity $H_p/v_{rms}$ varies from 
about a minute 
in the photosphere to 6 hours in 15.5 Mm depth. The effective life time of
convection patterns at the respective depths is about a factor of 5-10
larger, leading to time scales of 5-10 minutes in the photosphere (life time
of granules), 3-6 hours in 5.5~Mm depth (approximate life time of the sunspot
in Figure \ref{fig:f12}), and 30-60 hours in 15.5 Mm depth (approximate life 
time of the sunspot in Figure \ref{fig:f13}). From this we can conclude that
the the large scale flow component (3) can be considered mostly converged
in the upper most 8~Mm of our simulation domain (at least 5-10 convection
pattern life times). In addition we did not see any evidence that the temporal
evolution changes the large scale flow topology, it mostly affects the average
flow amplitudes (see Figure \ref{fig:f2}). As far as it concerns the flow 
component (3) the current simulations likely underestimate the flow velocity 
in the deeper parts of the domain.

Related to the overall temporal evolution is also the initial state of our 
simulations and its influence on the subsequent sunspot evolution as 
discussed in Section \ref{sect:bottom-bnd}. In our current setup large scale
flows are the response to a magnetic obstacle in the convection zone that is
initially not consistent with the flow fields present. However, the system 
evolves into a state in which both become consistent by evolving the 
magnetic field and flow structure. Ultimately this shortcoming will be resolved
through active region scale flux emergence simulations. Fist results such as
\citet{Cheung:etal:2010} and work in progress lead to results that are in 
terms of magnetic field and flow structure comparable to the penumbra free 
sunspot discussed here.

\subsection{Subsurface field structure of sunspots}
\label{sect:dis:field}
We investigated the influence from the bottom boundary on the overall 
stability and life time of our simulated sunspots by
comparing simulations in $6.144$ and $16.384$~Mm deep domains. 
In the shallow domain a closed boundary condition in 
strong field regions is required to prevent an unrealistically fast 
decay of the sunspot. Using the same boundary condition in the $16$~Mm deep
domain adds several degrees of freedom to the long term evolution of the
sunspot allowing for flux separation and decay driven through convective 
motions in more than $10$~Mm depth beneath the solar surface. In the photosphere
these flux separation events are accompanied with light bridge formation.
Changing the bottom boundary condition to an open boundary in strong
field regions enhances the deformation and decay rate of the sunspot, but
unlike shallow domains, the sunspot stays coherent for at least 24 hours
in time. It is very likely that the time scales of sunspot decay will become
consistent with observed sunspot life times in deeper computational domains
regardless of the bottom boundary condition used. 
We note that these conclusions are currently based on numerical simulations 
of sunspots without an extended penumbra. It has been suggested that the 
presence of an extended penumbra can lead in the upper most few Mm to an 
additional stabilization
of a sunspot against interchange instabilities \citep{Meyer:etal:1977}.
Note that our simulation with penumbra was performed only in a 9 Mm deep 
domain (due to the substantially higher grid resolution) and used again a 
closed boundary in strong field regions. 

The ``anchoring problem'' of sunspots \citep[see for example the corresponding 
chapters in][and further references therein]
{Gizon:etal:2010:review,Moradi:etal:2010} 
is closely related to the subsurface structure of sunspots. 
A sufficiently deep anchoring is required to explain the relatively long 
life times of sunspots compared to the convective turnover as well as 
Alfv{\'e}nic travel time scales found in the upper convection zone. The two
solutions discussed are the cluster model \citep{Parker:1979b}, in which a
convergent subsurface flow opposes sunspot decay, and a sufficiently deep
anchoring close to the base of the convection zone where convective turnover 
time scales (defined through $H_p/v_{rms}$) are of the order of 1-2 weeks or 
even longer (overshoot region). Note that this is about a factor of 30
longer than the longest time scales in our 16~Mm deep simulation
domain, in addition we found that the effective life time of the magnetic
structure is about $5-10\,H_p/v_{rms}$.

With regard to the cluster model we do not see clear evidence that converging
flows in the proximity of sunspots are a viable solution -- at least not
for a sunspot with an extended penumbra where we find outflows at all 
depth levels. Although a weaker converging flow is present in our
sunspot model without penumbra, we do not see a clear indication that it
stabilizes the sunspot sufficiently against decay (see the case in Figure
\ref{fig:f13}).

With regard to ``deep anchoring'' it has been pointed out that this might be
inconsistent with the photospheric motion of sunspots after the flux emergence
process. Rising magnetic flux tubes show typically rather long wavelengths such
as $m=1$ or $m=2$ modes \citep[c.f.][]{Fan:etal:1993,Fan:etal:1994,
Moreno-Insertis:etal:1994,Schuessler:etal:1994,Caligari:etal:1995}. 
Due to magnetic tension forces
the opposite polarities of the active region continue to separate after the
emergence at a rate that is inconsistent with observations. Because of this
it has been suggested that sunspots become dynamically disconnected from 
their magnetic roots (at the base of the convection zone) shortly after the 
flux emergence in the photosphere \citep{Fan:etal:1994,Moreno-Insertis:etal:1995,Schrijver:Title:1999,Schuessler:Rempel:2005}. Here the term ``dynamical 
disconnection'' is used to describe a situation in which the magnetic field
strength drops over a certain height range below equipartition field strength
so that magnetic field becomes passive with respect to convective motions.
While the dynamical disconnection process 
alleviates the drift problem mentioned above, there are currently 2 unresolved
issues: 1. Is it possible to dynamically disconnect a flux tube as suggested
by \citet{Schuessler:Rempel:2005} without destroying the coherence of the
sunspot in the photosphere? 2. Is the resulting shallow (possibly less
than $10$~Mm deep) sunspot sufficiently stable?

The models presented here give some clues to answer these questions.
In the simulation with open bottom boundary condition in a $16$~Mm domain
we observe initially
(after switching from a closed to an open boundary) an inflow into the
sunspot with an amplitude comparable to the typical convective rms velocity 
at the depth of the boundary condition, i.e. a velocity of a few 
$100\,\mbox{m\,s}^{-1}$. The latter is very similar to the assumptions made 
by \citet{Schuessler:Rempel:2005}.
The inflow leads to a destruction of the coherent trunk of the 
spot in the lower part of the domain while the field in the photosphere shows
only a moderate enhancement of decay compared to the closed boundary reference 
case. While this is in principal support of the disconnection scenario 
discussed by \citet{Schuessler:Rempel:2005}, our results point also toward
the necessity of a sufficiently deep disconnection to ensure coherence of 
the photospheric parts of the sunspot over time scales of several days to
weeks. In our currently $16$~Mm deep domain we observe a slow decay of the 
sunspot on time scales of days. Extrapolating this result, life times of about 
a week should be reached in a depth of about $30-50$~Mm. In addition it is
conceivable that the subsurface convection pattern is substantially altered
as consequence of the flux emergence process forming the sunspots in the first
place. This could lead to systematic difference compared to the simulations 
presented here, in which we added the magnetic field to a pre-existing 
convection pattern.
This problem will be ultimately resolved once realistic flux emergence 
simulations such as \citet{Cheung:etal:2010} will become available in 
sufficiently deep domains, perhaps coupled with models of flux emergence
in the deep convection zone such as \citet{Fan:2008}.  

With regard to the subsurface structure we point out that the numerical
simulations presented here and especially those described in 
\citet{Rempel:etal:Science} are clearly biased toward the monolithic
picture due to the monolithic initial state as well as bottom boundary
condition (at least if the closed boundary is used). Nevertheless, we
find in the deeper domains considered here a substantial fragmentation of
the subsurface field caused by the interaction with the surrounding
convection. Almost all fragmentation events (intrusions of field 
free plasma) present several Mm beneath the surface lead on time scales of 
a few hours to a day to the formation of light bridges or flux separation in 
the photosphere. These results allow for the conclusion that the subsurface 
field of sunspots very likely shows significant fragmentation, but most of
these subsurface fragmentations (at least those present in the upper most 
$10$~Mm) do not remain hidden and become visible in the photosphere 
on a rather short time scale, i.e. the photospheric appearance of a sunspot 
should tell us a lot about its subsurface structure: Sunspots that are very 
stable and do not show light bridges are more monolithic than sunspots with 
light bridges and signs of flux separation.

\section{Conclusions}
\label{sect:conclusions}
We presented simulations of sunspots in up to $16.384$ Mm deep and up to
$73.728$ Mm wide domains covering time evolution of up to 2 days. Through
variations of the magnetic top boundary condition we simulated sunspots
with and without penumbrae while having comparable magnetic flux, allowing
for a direct side by side comparison of their properties with regard to large
scale flows. We identified 5 different flow components in our simulations,
which we discussed in Section \ref{sect:dis:flow} and highlighted possible
connections to the Evershed - moat flow connection in Section 
\ref{sect:dis:evershed_moat}. Overall we identified in our simulations 2
outflow components (one shallow and one deep reaching) that give a 
contribution to large scale flows around sunspots independent from the 
Evershed flow. In the case of a sunspot with penumbra we find outflows at all
depths covered by the numerical simulation. Resolving the Evershed moat flow 
connection requires in addition to the detailed study of photospheric flows 
also the study of deeper reaching flow structures that are accessible 
through helioseismology.

The long-term stability and evolution of the simulated sunspots is mostly
governed by the longest time scale of convective motions found near the bottom 
of the simulation domain. If the magnetic field is not artificially 
constrained at the bottom boundary, sunspots strongly deform and decay on
a time scale of the order of $10\,H_p/v_{rms}$ (evaluated near the bottom of 
the domain). This translates to a few hours in 6~Mm depth, about 1-2 days 
in 16~Mm depth, and 10~days in 50~Mm depth (assuming the trend indicated in 
the numerical simulations shown here is valid over a broader depth range).
We did not find a clear indication that the large scale flows developing
around our simulated sunspots play a key role in maintaining their coherence,
at least not beyond the intrinsic convective time scales present. In that
sense our results point more toward ``deep anchoring'' as an explanation
for the observed sunspot life times. Further clarification of this matter
will require simulations in deeper domain that also include the flux emergence
and sunspot formation process in order to start from a more consistent 
initial state.

\acknowledgements
M. Rempel is partially supported through NASA grant NNH09AK02I (SDO Science 
Center) to the National Center for Atmospheric Research. NCAR is sponsored 
by the National Science Foundation. The author thanks Y. Fan, A. Birch, D.
Braun, V. Mart{\'i}nez Pillet for fruitful discussion and comments on the
manuscript, and the anonymous referee for helpful suggestions. 
This work was made possible by NASA's High-End Computing Program as well as
NSF computing resources provided through the Teragrid. The simulations 
presented in this paper were carried out on the Pleiades cluster 
at the Ames Research Center under project GID s0925, the Texas Advanced 
Computing Center (TACC) under grant TG-MCA93S005 as well as the National 
Institute for Computer Sciences (NICS) under grants TG-MCA93S005 and 
TG-AST100005. We thank the staff at the supercomputing centers for their 
technical support. 

\bibliographystyle{natbib/apj}
\bibliography{natbib/papref_m} 

\end{document}